\documentclass[conference]{IEEEtran}
\usepackage{url,bm,algorithm,algorithmic,epsfig,endnotes}
\usepackage{mathrsfs}                                                  
\usepackage{balance}
\usepackage{amsmath,graphics,epstopdf,color,enumerate,subfigure}
\graphicspath{{figure/}}
\newtheorem{mythe}{Theorem}

\newtheorem{mydef}{Definition}

\usepackage{caption}
\usepackage{makecell}
\usepackage{hyperref}
\usepackage{balance}
\usepackage{courier}
\usepackage{amsfonts}

\newcommand{\blue}{\textcolor{black}}
\newcommand{\ccs}{\textcolor{black}}
\newcommand{\ccsnew}{\textcolor{black}}
\newcommand{\ccsfinal}{\textcolor{black}}

\newcommand{\chang}{\textcolor{black}}
\newcommand{\needcheck}{\textcolor{black}}
\newcommand{\unknown}{\textcolor{black}}
\newcommand{\revise}{\textcolor{black}}
\newcommand{\secondrevise}{\textcolor{black}}
\newcommand{\usenix}{\textcolor{black}}
\newcommand{\thirdrevise}{\textcolor{black}}
\newcommand{\ndss}{\textcolor{black}}
\newcommand{\ndsstwo}{\textcolor{black}}
\newcommand{\ndssthree}{\textcolor{black}}

\newcommand{\SP}{\textcolor{black}}

\newcommand{\tabincell}[2]{\begin{tabular}{@{}#1@{}}#2\end{tabular}}%
\makeatletter
\def\@copyrightspace{\relax}
\makeatother

\begin{document}

\title{DEEProtect: Enabling \ccs{Inference Control} on Mobile Sensing Applications} 
\author{Changchang Liu$^1$~~~~~~Supriyo Chakraborty$^2$~~~~~~Prateek Mittal$^1$\\
$^1$ Princeton University, $^2$ IBM T.J. Watson Research Center}
\maketitle

\begin{abstract}
Personal sensory data is used by context-aware mobile applications to provide utility. However, the same data can also be used by an adversary to make sensitive inferences about a user thereby violating her privacy. 

We present DEEProtect, a framework that enables a novel form of \emph{inference control}, in which mobile apps with access to sensor data are limited (provably) in their ability to make inferences about user's sensitive data and behavior. DEEProtect adopts a two-layered privacy strategy. First, it leverages novel autoencoder techniques to perform data minimization and limits the amount of information being shared; the learning network is used to derive a compact representation of sensor data consisting only of features relevant to authorized utility-providing inferences.
Second, DEEProtect obfuscates the previously learnt features, thereby providing an additional layer of protection against sensitive inferences. Our framework supports both conventional as well as a novel relaxed notion of local differential privacy that enhances utility. Through theoretical analysis and extensive experiments using \SP{real-world datasets}, we demonstrate that when compared to existing approaches DEEProtect provides provable privacy guarantees with up to $8x$ improvement in utility. Finally, DEEProtect shares obfuscated but raw sensor data reconstructed from the perturbed features, thus requiring no changes to the existing app interfaces.
\end{abstract}

%
%

\section{Introduction}
\label{sec:intro}
Mobile devices such as smartphones and wearable devices are increasingly gaining popularity as platforms for collecting and sharing sensor data. These devices, often equipped with sensors such as accelerometer, orientation sensor, magnetometer, camera, microphone, GPS and so on, are being used by mobile sensing systems to make sophisticated inferences about users. These inferences have in turn enabled an entire ecosystem of context-aware apps such as behavior-based authentication \cite{frank2013touchalytics,lee2015multi,mare2014zebra,conti2011swing,riva2012progressive,zhu2013sensec}, fitness monitoring based on activity tracking~\cite{bao2004activity,reddy2010using,luxton2011mhealth}, speech translation~\cite{lei2013accurate,michalevsky2014gyrophone} and traffic/environmental monitoring~\cite{azizyan2009surroundsense,mohan2008nericell,PlaceRaider,tung2015echotag}.

While shared data has enabled context-aware apps to provide utility to the users, the same data can also be used by an adversary to make inferences that are possibly sensitive to the user such as speaker identity~\cite{liu2012cloud,nirjon2013auditeur}, keystroke detection~\cite{xu2012taplogger,liu2015good,cai2011touchlogger,marquardt2011sp,mohamed2017smashed}, location tracking~\cite{brouwers2011detecting,kim2010sensloc,nirjon2013auditeur}, device placement~\cite{park2012online}, onscreen taps recognition~\cite{miluzzo2012tapprints}, onset of stress~\cite{lu2012stresssense,chang2011ammon} and detection of emotional state~\cite{chang2011ammon,rachuri2010emotionsense}. 
{Therefore, there exist fundamentally conflicting requirements between protecting privacy of the sensitive information contained in mobile sensor data and preserving utility of the same data for authorized inferences.} We consider the following case studies that further illustrate this privacy-utility tradeoff:
\begin{enumerate}[$\bullet$]
\item {Case Study 1}: \chang{A user} shares accelerometer data with an authentication app that performs keyless validation of user identity (\emph{useful inference})~\cite{frank2013touchalytics,lee2015multi,mare2014zebra}. But the same data can also be used to infer the activity mode of the user -- if she is walking, or standing still, or moving up or down the stairs (\emph{sensitive inferences})~\cite{bao2004activity,reddy2010using,luxton2011mhealth} -- which in turn may lead to more serious inferences regarding living habits and health conditions.
\item {Case Study 2:} \chang{A user} shares accelerometer data with a real-time life-logging app that helps with travel planning (\emph{useful inferences})~\cite{kim2010sensloc,han2012accomplice}. However, the same data can also be maliciously used to extract sequences of the user's entered text (e.g., Passwords or PINs) (\emph{sensitive inference})~\cite{xu2012taplogger,miluzzo2012tapprints,cai2011touchlogger,marquardt2011sp}. 
\item {Case Study 3:} \chang{A user} shares speech data with a voice-based search app that translates speech to text for web searches (\emph{useful inference})~\cite{lei2013accurate,michalevsky2014gyrophone}. But the same data can also be used to recognize the user's identity (\emph{sensitive inference})~\cite{liu2012cloud,nirjon2013auditeur}, compromising her privacy.
\end{enumerate}

Recently, several sensor privacy protection mechanisms (SPPMs) have been proposed~\cite{beresford2011mockdroid,hornyack2011these,cornelius2008anonysense,shebaro2014identidroid,de2011short,de2012preserving, li2013efficient,gotz2012maskit, chakraborty2014ipshield, boxify}. However, most existing SPPMs only provide binary access control over the sensor data that requires users to choose between sharing or blocking a sensor -- limiting their applicability and adoption~\cite{gotz2012maskit,shebaro2014identidroid,beresford2011mockdroid,hornyack2011these}. A limited set of SPPMs that do focus on obfuscating sensor data, however, \chang{lack} provable privacy guarantees~\cite{chakraborty2014ipshield,boxify}. 

\begin{figure}[!t] \centering
\includegraphics[width=2.5in,height=1.3in]{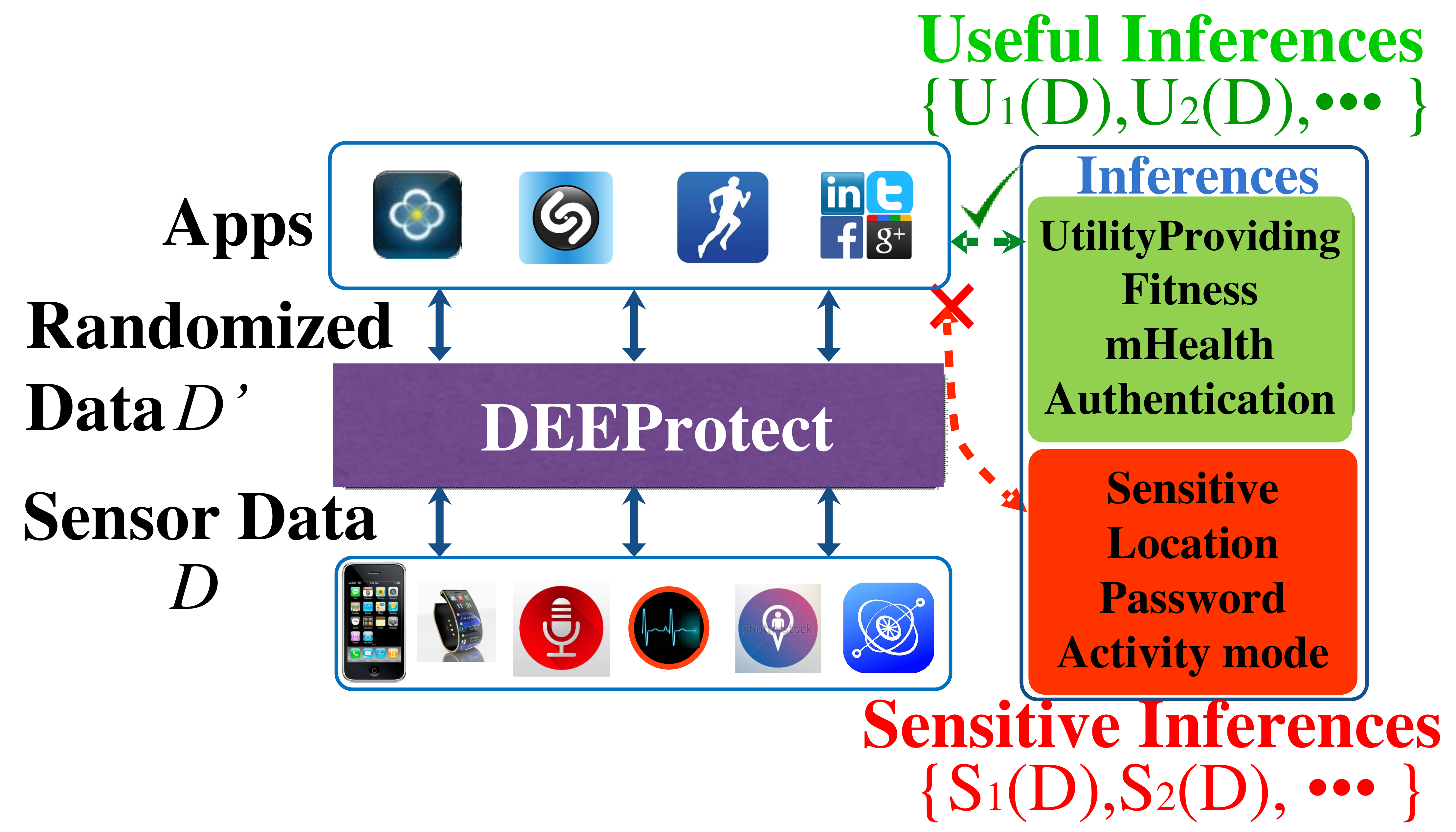}
\vspace{-0.8em}
\caption{{DEEProtect mediates privacy-preserving 
access to mobile sensor data \ndss{through inference control}. Mobile apps can access obfuscated sensor 
data for providing utility, but are provably limited in their ability to
infer sensitive information.}}
\label{fig:motivation}
\vspace{-1em}
\end{figure}

In this paper, we propose a new framework called DEEProtect that mediates privacy-preserving access to mobile sensor data. {DEEProtect enables a new permission model where instead of blocking or allowing sensors, \ndss{the system} specifies the utility and privacy preferences in terms of inferences that can be derived from mobile sensor data.} The overall information flow of DEEProtect is summarized in Figure~\ref{fig:motivation}. \ndss{To the best of our knowledge, we are the first to propose a rigorous approach for inference control on smartphones.}

\ccsnew{In DEEProtect, \ndss{the system configuration} can specify as input a set of useful/authorized inferences denoted by ${U}(D)=\{{U}_1({D}),{U}_2({D}),\cdots\}$, e.g., behavior-based authentication, speech-to-text translation. All other inferences are considered sensitive by default. Alternatively, a subset of inferences can be specified as being sensitive, which is denoted by ${S}(D)=\{{S}_1({D}),{S}_2({D}),\cdots\}$, e.g., detection of the text entered in keyboard, speaker identity recognition. Note that we provide a list of possible inferences as input interface of our system. In our setting, inferences represent functions that map the mobile sensor data to their corresponding inference results (i.e., labels).} \ndss{These functions could vary in complexity from linear to highly non-linear models.} Outgoing sensor data is initially intercepted by DEEProtect, which obfuscates the data before sharing such that ${U}(D)$ can be derived accurately by the app and ${S}(D)$ is kept private.

\ccsnew{To provide provable privacy guarantees for a single user's mobile sensor stream, we exploit \emph{local differential privacy}~\cite{duchi2013local,erlingsson2014rappor,qin2016heavy} to obfuscate the mobile sensor data--which unlike differential privacy~\cite{dwork:Springer06,dwork:SpringerTC06,dwork2008differential,dwork2010differential,dwork:ACMC11} \ndss{\emph{does not require a multi-user setting}}. However, adding noise directly to the high-dimensional data reduces its utility significantly.}

{Towards this end, we propose a computationally efficient mechanism that strikes an enhanced balance between utility and privacy guarantees. We start with a data minimization step, in which we leverage a novel proposal of a supervised autoencoder to extract features that are relevant only to the useful inferences. \SP{The reduced dimensionality of the data helps improve the overall system utility}. Then, we perturb these extracted features, using either the conventional local differential privacy framework~\cite{duchi2013local,erlingsson2014rappor,qin2016heavy} or our \emph{relaxed} notion of {local differential privacy}, depending on the specification of sensitive inferences. Finally, we reconstruct obfuscated sensor data from the perturbed features such that the existing interfaces for mobile apps can continue to work without any change.} \ndsstwo{Note that our approach is of independent interest even outside the context of smartphone applications. For example, our algorithm can be applied to enforce general data protection regulation (GDPR) in the European Union laws, in the context of protecting user privacy.}

\noindent{\bf Contributions:} We summarize our contributions as follows:
\begin{enumerate}[$\bullet$]
\item {\revise{\em Provable Privacy Guarantees in Single-user Settings:}} {We support rigorous local differential privacy guarantees for protecting \ndss{\emph{a single user's mobile sensor data}.} \ndss{Furthermore, we propose} a novel relaxed variant of local differential privacy for providing \ccs{inference control} over the mobile sensor data, \ndss{which is the first such attempt to the best of our knowledge.} \secondrevise{Our proposed privacy metric satisfies composition properties, and can achieve improved utility by focusing on protecting a specific set of inferences.}} 

\item {\revise{\em Novel Perturbation Mechanisms for Enhanced Privacy-Utility Tradeoffs:}} \revise{We propose an effective perturbation mechanism which consists of \ccs{two key techniques: (1)} \ccs{autoencoder} based data minimization and \ccs{(2)} feature obfuscation based data perturbation. Our first technique automatically extracts features relevant to the useful/authorized inferences by extending the \ccs{autoencoder} models for dimensionality reduction, which to our best knowledge is the first such attempt. For the second technique, we propose effective feature perturbation methods, that achieve both conventional and relaxed notions of local differential privacy guarantees. Through rigorous theoretical analysis, we demonstrate the advantage of our approach over the state-of-the-art obfuscation mechanisms.} 

\item {\em Implementation and Evaluation:} We implement our system using real-world datasets and mobile apps (for all the case studies discussed previously). For our first feature learning step, we demonstrate that our modified \ccs{autoencoder} network outperforms \SP{the state-of-the-art feature extraction approaches such as discrete \emph{Fourier} transform (DFT), discrete cosine transform (DCT), wavelet transformation, principal component analysis (PCA) and blind compressive sensing (BCS),} with up to $2x$ improvement. For our end-to-end approach combining both feature extraction and perturbation, we demonstrate the advantage of our technique over the state-of-the-art obfuscation mechanisms with up to $8x$ improvement. \ndss{Furthermore, we will make our DEEProtect framework publicly available.}

\end{enumerate}

\section{Related Work}
\label{related}
\subsection{Android-based SPPMs}
\usenix{MockDroid~\cite{beresford2011mockdroid} is a modified version of the Android operating system (OS) with ability to `mock' a given resource requested by an app. This approach allows users to revoke access to particular
resources at run-time, encouraging users to consider
the trade-off between functionality and privacy.
AppFence~\cite{hornyack2011these} offers two approaches for protecting sensitive data from untrusted apps: shadowing and blocking sensitive data from being exfiltrated off the device. IdentiDroid~\cite{shebaro2014identidroid} is also a customized Android OS providing different identifications and privileges, to limit the uniqueness of device and user information. The above systems still rely on the user to determine the sensitive sensors and block, mock, or shadow them. {ipShield~\cite{chakraborty2014ipshield}, provides users the ability to specify their privacy preferences in terms of inferences, however, they
 only generate binary privacy policies of \emph{allow} and \emph{deny} for individual sensors. Such binary access control over the sensor data may seriously affect the functionality of various apps. DEEProtect on the other hand provides the much needed automatic translation from higher-level privacy specifications in terms of inferences to obfuscation of sensor data.}} 
 
Another important limitation of the state-of-the-art SPPMs is that they are often heuristic in nature and fail to provide rigorous privacy guarantees for \ccs{inference control} over mobile sensors \cite{chakraborty2014ipshield,xu2015semadroid,sikder20176thsense,aware2017,petracca2015audroid}. For instance, ipShield~\cite{chakraborty2014ipshield} does provide users with ability to configure obfuscation policies for sensors (through addition of random noise), but does not quantify any privacy guarantees for those policies. Boxify~\cite{boxify} provides app sandboxing for untrusted apps in an isolated environment with minimum access privileges. EaseDroid~\cite{easedroid} uses semi-supervised learning to perform automatic policy analysis and refinement. This could be used to compliment DEEProtect by automatically learning the set of inferences that need to be protected and ones that need to be released from access patterns and audit logs.

\ccsnew{Suman et al. in \cite{jana2013enabling, jana2013scanner} presented novel approaches to add a privacy-protection layer to mobile systems where untrusted applications have access to camera resources. These methods primarily aim to limit the information contained in an image before sharing, and could potentially be used in conjunction with our DEEProtect -- e.g., the analysis of DEEProtect could inform transformations applied by Suman et al.}

Privacy-preserving mobile data aggregation has been studied in~\cite{ li2013efficient,de2012preserving,cornelius2008anonysense}. \ndss{These approaches assume the presence of a multi-user database and aim to protect the privacy of the multi-user training data. However, DEEProtect aims to protect a single user's data while using his/her mobile device. The training process in DEEProtect is implemented offline and only the trained models/parameters are transferred to the mobile device. Therefore, a privacy-preserving training process in~\cite{ li2013efficient,de2012preserving,cornelius2008anonysense} is \emph{orthogonal} to our scenario.} 
\subsection{Repository of Inferences}
We surveyed more than $70$ research papers published in relevant conferences and journals over the past $6$ years to form a knowledge repository of inferences \SP{(as shown in Table~\ref{tableknowledge}) in the Appendix} that can be made using a combination of sensors accessed by mobile apps. \ccsnew{In Table~\ref{tableknowledge}, each row represents a category of inference computed over mobile sensor data, along with the specific sensors that are utilized and the corresponding research papers.} \usenix{This table forms the universe of possible inferences over which the set of useful and sensitive inferences in DEEProtect are defined.} 
\subsection{\revise{Summary of Our Novelty}}
\secondrevise{To the best of our knowledge, DEEProtect is the first system that provides a novel \ccs{inference control} in mobile sensing applications. Furthermore, DEEProtect 1) \SP{uniquely} provides provable privacy guarantees including both conventional local differential privacy and our relaxed variant of local differential privacy for sensitive sensor data; 2) presents an effective mechanism consisting of two key techniques: \ccs{autoencoder} based data minimization and feature obfuscation based data perturbation; 3) outperforms the state-of-the-art methods in multiple mobile sensing datasets and applications.}

\section{Threat Model and System Overview}
\label{systemdesign}
\ccsnew{In this section, we will discuss the threat model and the two usage modes of our DEEProtect system. We will also provide an overview of our approach. }
\subsection{\ccs{Threat Model and Usage Modes}}
\label{sec_threat}
We assume that DEEProtect together with the underlying mobile OS, sensors, and system services form the trusted domain. Our adversary is an untrusted app provider, who publishes apps in the marketplace for advertised useful inferences. The app accesses sensors including ones to which the user has provided explicit permission, and also others for which no permission is required. \usenix{The app would then send out the collected sensor data to the adversary.} Our goal is to ensure that the data shared with the app can only be used for deriving the authorized (useful) inferences and not any sensitive inferences. {We therefore aim to limit/bound the adversary's inference of the user's sensitive information \SP{by mediating access to obfuscated sensor data provided by DEEProtect} while achieving rigorous privacy properties. Specifically, DEEProtect supports two usage modes, depending on how sensitive inferences are specified.}

\ccs{\emph{Usage Mode 1:} Only the useful inference set ${U}(D)$ is provided and the sensitive inference set $S(D)$ is the default set of `all possible inferences'. Under this mode, the user is interested in rigorous privacy guarantees with respect to the entire set (possibly unknown) of inferences. \SP{DEEProtect under usage mode 1 aim to provide rigorous local differential privacy for mobile sensor data.}}

\ccs{\emph{Usage Mode 2:} Both the useful and the sensitive inference sets are provided. Under this scenario, the user is interested in protecting only a specific set of sensitive inferences (and not all possible inferences). \SP{To provide provable privacy guarantees, we develop a relaxed variant of {local differential privacy} and use it to perturb the mobile sensor data.}}

\begin{figure}[!t] \centering
\subfigure[]{
\label{venn_diagram} 
\includegraphics[width=2in,height=0.8in]{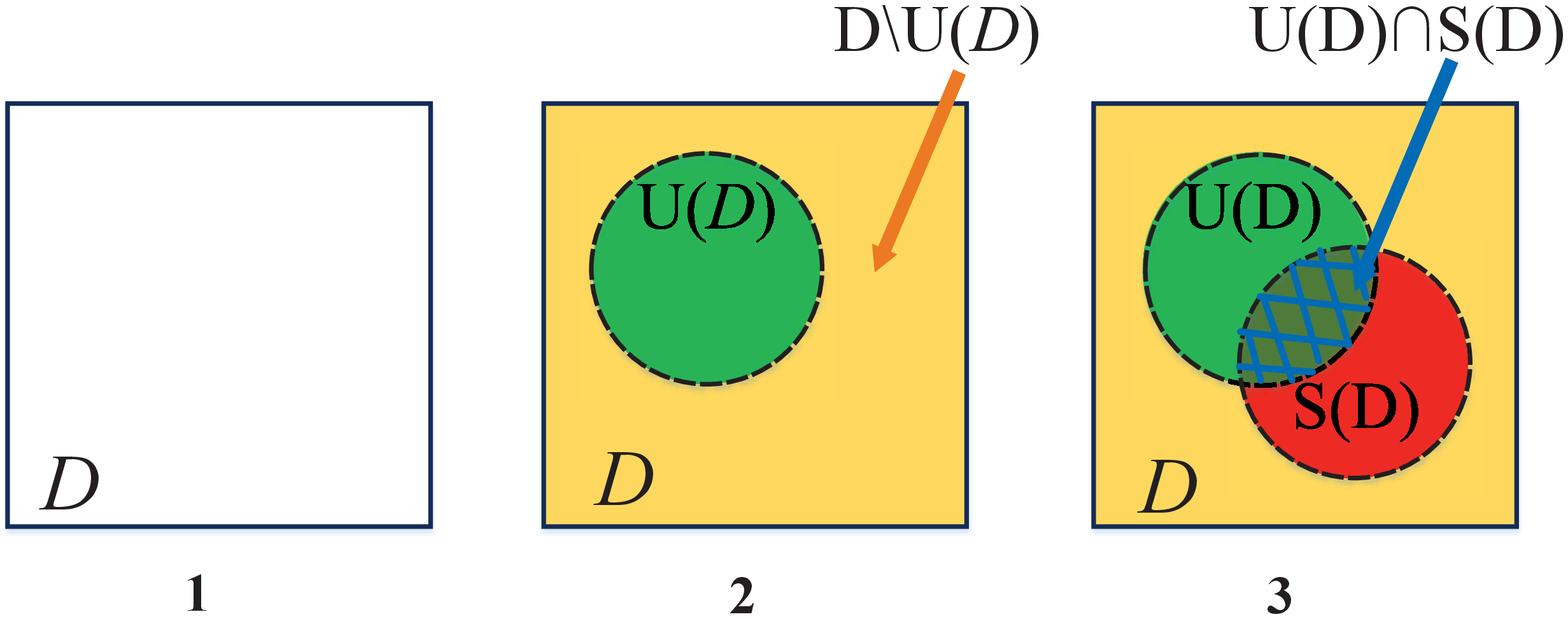}}
\hspace{-0.1in}
\subfigure[]{
\label{fig:pipeline} 
\includegraphics[width=1.3in,height=0.8in]{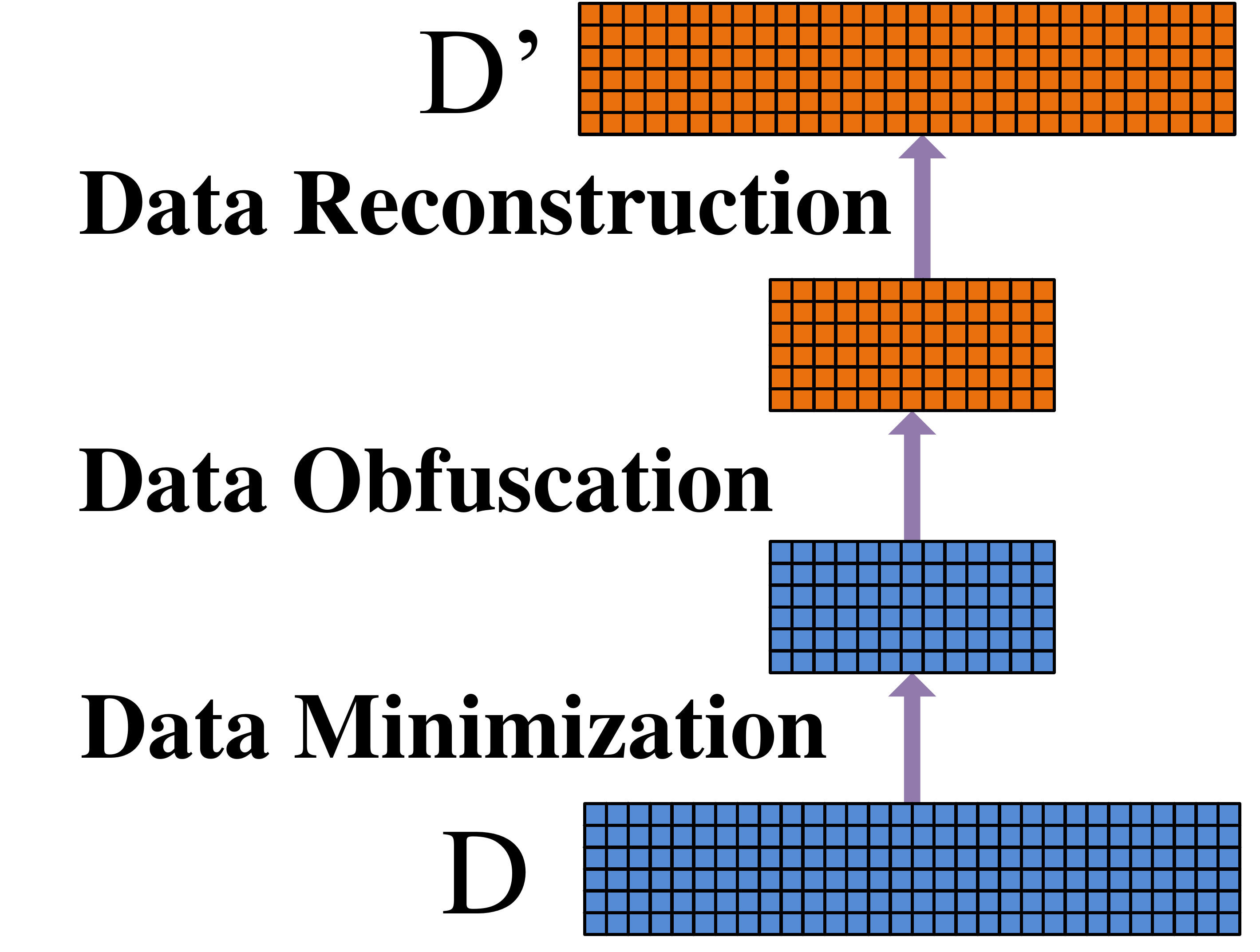}}
\vspace{-1em}
\caption{\chang{(a) Venn diagram for information measures associated with user's raw sensor data $D$ (shown in white), useful inferences ${U}(D)$ (shown in green) and sensitive inferences ${S}(D)$ (shown in red). (b) DEEProtect pipeline comprises of a data minimization step for learning a compact feature representation specific to authorized useful inferences; a data obfuscation step to perturb features and provide provable privacy guarantees; a data reconstruction step to generate obfuscated sensor data from perturbed features.}}
\label{fig:deeprotect}
\vspace{-1em}
\end{figure}

{It is also interesting to note that a threat model similar to usage mode 2 (i.e., protecting against specified inferences) has been explored in Pufferfish privacy~\cite{kifer2014pufferfish} and Blowfish privacy~\cite{he2014blowfish}. For example, the set of sensitive inferences computed over sensor data in DEEProtect is the set of potential secrets defined in Pufferfish and Blowfish privacy. \usenix{Blowfish privacy has recently been adopted} by the {U.S. Census Bureau} demonstrating \ndss{the usefulness of this paradigm for enabling practical deployment.}}

\subsection{\ccs{Approach Overview}}\label{sec_approach}
\ccsnew{We now illustrate the working of DEEProtect by using Case Study 3 as an example.  Figure~\ref{venn_diagram} shows the venn diagram of information measures in DEEProtect. \ccsnew{To protect users' privacy, it is required that we should not directly share the raw sensor data $D$ with the untrusted third-party apps, where $D$ is the entire white square shown in Figure~\ref{venn_diagram}--1.} In Case Study 3, $D$ corresponds to the microphone data. Our objective is to enable $U(D)$ ({speech-to-text translation}) with high accuracy while protecting against $S(D)$ ({speaker identity recognition}).} 

{\SP{\bf{Intuition of Our Key Idea Using Case Study 3:}}} The key idea in DEEProtect is to first ignore the information that is orthogonal to the useful inference information, \SP{since the overall utility performance can not benefit from these orthogonal information.} In this case, we only retain information that enable ${U}(D)$ and ignore the remaining information ${U}^c(D)=D\backslash {U}(D)$, which is shown in yellow in Figure~\ref{venn_diagram}--2. In the next step, we perturb those information that are correlated to both the useful and sensitive inferences, ${U}(D)\cap {S}(D)$, as shown in blue in Figure~\ref{venn_diagram}--3. \SP{The underlying reason is that these correlated features can help the adversary to infer sensitive information of the users.  This feature perturbation step helps in selectively hindering the sensitive inferences without severely degrading the useful inferences, while providing rigorous local differential privacy under usage mode 1 and relaxed location differential privacy under usage mode 2.}

\SP{\bf{Description of Our Approach Using Case Study 3:}} {Intuitively, both the useful and sensitive inferences rely on a limited number of features extracted from the sensor data. The commonly used features for {speech-to-text translation} are {Mel-frequency Cepstral Coefficients (MFCC)}, {spectrogram} and {N-gram}~\cite{lei2013accurate,michalevsky2014gyrophone,zhou2013ibm}, while the popular features for {speaker identity recognition} are {MFCC}, {spectrogram} and {voice biometrics} of the user (such as her pitch, accent, etc.)~\cite{liu2012cloud,nirjon2013auditeur}. {The first challenge for DEEProtect is to learn the set of features associated with the useful/authorized inferences.} To mitigate this problem, DEEProtect modifies autoencoder models~\cite{hinton2006reducing,lee2008sparse,vincent2008extracting} by explicitly incorporating the useful inferences to automatically learn the appropriate set of features pertaining to the useful inferences. {This implies that we only retain {MFCC}, {spectrogram} and {N-gram} features that are required for speech-to-text translation\footnote{Note, these automatically learned features can not be interpreted in the same way as MFCC, spectrogram, N-gram, etc. For ease of explanation, we continue to name them as MFCC, spectrogram, N-gram, etc.}. In the next step, we perturb the MFCC and spectrogram features, which are required for both speech-to-text translation and speaker identity recognition. Finally, we reconstruct the privacy-preserving microphone data from the perturbed MFCC and spectrogram, and the unperturbed N-gram before sharing with the third-party apps.}}

\SP{\bf{Approach Overview for Arbitrary Inferences:}} For arbitrary useful and sensitive inferences, the three-stage data obfuscation pipeline used in DEEProtect is illustrated in Figure~\ref{fig:pipeline}. In the first stage, we extract features from raw sensor data and perform \emph{data minimization}~\cite{house2012consumer,pfitzmann2010terminology,cavoukian2009privacy} to guarantee that only features that are relevant to the authorized (useful) inferences are retained for further processing. Specifically, DEEProtect modifies an autoencoder network~\cite{hinton2006reducing,lee2008sparse,vincent2008extracting} to \emph{automatically} explore the inherent structural characteristics of the sensor data, and learn a sparse feature representation of the data specific to the useful inferences. At this point, some of the extracted features, although highly specific to the useful inferences, might still have some correlation with the sensitive inferences. \ccs{In the next step of the pipeline, we develop a computationally efficient mechanism that perturbs the sparse features, using either the conventional local differential privacy framework~\cite{duchi2013local,erlingsson2014rappor,qin2016heavy} or our relaxed notion of {local differential privacy} framework, depending on the specification of sensitive inferences. Finally, we reconstruct the obfuscated sensor data from the perturbed features such that the existing interfaces for mobile apps can continue to work without any change.} \ccs{Specifically, we adapt the popular Laplace perturbation mechanism~\cite{dwork:Springer06} to construct effective perturbation approaches, in order to provide both conventional and relaxed notions of local differential privacy guarantees for protecting mobile sensor data. {Our theoretical analysis and experimental results on real-world datasets show that both usage modes of our approach significantly outperform previous \thirdrevise{state-of-the-art} with up to 8x improvement.}}

\section{Privacy Metrics for Protecting Mobile Sensor Data in Single-user Settings}
\label{sec_perturbation}
\ccsnew{To provide provable privacy guarantees for a single user's mobile sensor stream, we exploit \emph{local differential privacy} to protect against all sensitive inferences (corresponding to usage mode 1 in Section~\ref{sec_threat}). We also propose a relaxed variant of local differential privacy to protect against a specific set of sensitive inferences (corresponding to usage mode 2 in Section~\ref{sec_threat}). We further demonstrate the privacy properties of our new privacy metric.}
\subsection{\revise{Local Differential Privacy}}
\label{sec:DP}
\revise{Differential privacy (DP)~\cite{dwork:Springer06,dwork:SpringerTC06,dwork2008differential,dwork2010differential,dwork:ACMC11} is a rigorous mathematical framework that prevents an attacker from inferring the presence or absence of a particular record in a statistical database. DP randomizes the query results, computed over the \emph{multi-user} database, to ensure that the risk to an individual record's privacy                               
does not increase substantially (bounded by a function of the privacy budget $\epsilon$) as a result of participating in the database.  Local differential privacy (LDP)~\cite{duchi2013local,erlingsson2014rappor,qin2016heavy}, as an \ndss{adaptation} of DP, is defined under the setting where the user does not trust anyone (not even the central data collector). Local privacy dates back to Warner~\cite{warner1965randomized}, who proposed the randomized response method to provide plausible deniability for individuals responding to sensitive surveys. In our setting, we therefore apply LDP to protect a \emph{single} user's mobile sensor data while satisfying rigorous privacy guarantees. Similar applications have also been explored in Google's Rappor System~\cite{erlingsson2014rappor}. Specifically, $\epsilon$-LDP is defined as follows:}
\begin{figure}[!t]
	\centering
	\includegraphics[width=2.7in,height=1.3in]{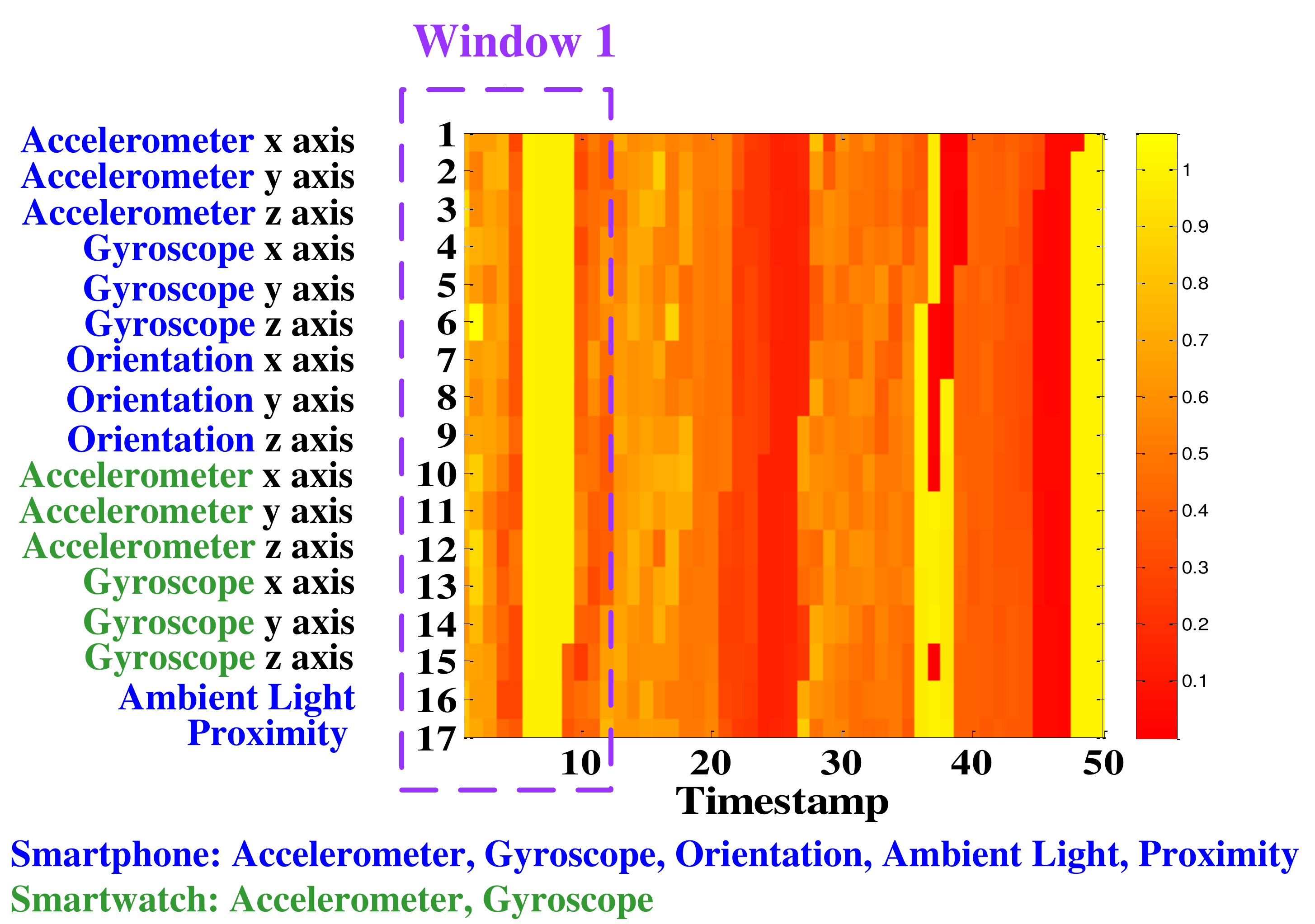}
	\DeclareGraphicsExtensions.
	\caption{Our collected mobile sensor data \secondrevise{(segmented with a time window size $N_w=10$)}.}
	\label{twodata}
	\vspace{-1em}
	\end{figure} 
\begin{mydef}\label{dpdef} {\emph{($\epsilon$-LDP)~\cite{duchi2013local}}}
 A randomized algorithm ${A}(\cdot)$ provides $\epsilon$-LDP if for \usenix{any two databases} ${D}_1, {D}_2$ and any output set ${o}$,
\begin{equation}\label{DP}
\begin{aligned}
\max_{{D}_1, {D}_2,o}\frac{\mathbb{P}({A}({D}_1)={o})}{\mathbb{P}({A}({D}_2)={o})}\le\exp(\epsilon)
\end{aligned}
\end{equation}
where ${A}({D}_1)$ (resp.${A}({D}_2)$) is the output of ${A}(\cdot)$ on input ${D}_1$ (resp.${D}_2$) and $\epsilon$ is the privacy budget. Smaller value of $\epsilon$ corresponds to a higher privacy level.
\end{mydef}	

{\bf{Applying LDP to Mobile Sensor Data:}} \usenix{For a mobile sensing system, consisting of $N_s$ sensors across $T$ timestamps, the temporal mobile sensor matrix $\bm{TMS}\in\mathbb{R}^{N_s\times T}$ is a real matrix where $\bm{TMS}(i,j)$ records the sensing data of the $i$-th sensor at the $j$-th timestamp. To fully explore the temporal characteristics of the data, we follow common practice~\cite{frank2013touchalytics,lee2015multi,luo2009compressive} and partition it into a series of segments according to a time window size $N_w$ \footnote{\ccs{We set window size as an input to our system.}}{as shown in Figure~\ref{twodata}}. For the $w$-th window, we stack the corresponding columns within it to form a column vector which is denoted by ${\bm{x}}_t \in \mathbb{R}^{N_sN_{w}\times 1}$. The temporal mobile sensor matrix can thus be reformulated as $\bm{X}=[{\bm{x}}_1; {\bm{x}}_2; \cdots]$, where $\bm{X}\in \mathbb{R}^{N_sN_w\times \frac{T}{N_w}}$. \revise{We therefore apply LDP to a single user's mobile sensor data which is segmented according to a window size. In this way, we can take the temporal dynamics of users' mobile sensor data into consideration, while providing rigorous privacy guarantees.}}

\revise{From Definition~\ref{dpdef}, we observe that $\epsilon$-LDP has the same mathematical formulation as $\epsilon$-DP, except that the neighboring databases in LDP are \emph{any two possible databases without the constraint of one tuple's difference as in DP~\cite{dwork:Springer06}}. Thus, when applying LDP to mobile sensing applications, the neighboring databases  represent any two segmented sensor data that differ in \emph{any possible sensor recording at any timestamp within the same window}. A mechanism that provides rigorous LDP can defend against any inference attacks over the obfuscated data, according to the post-processing invariant property~\cite{dwork2014algorithmic}. Therefore, we can provide rigorous LDP for users' sensor data to defend against all possible inferences over the data (corresponding to usage mode 1 in Section~\ref{sec_threat}).}

\subsection{\revise{Relaxed Local Differential Privacy}}
\label{sec:relaxdp}
\revise{{\bf{The need for a relaxed variant:}} Although we can provide rigorous LDP and defend against all inferences computed over the obfuscated sensor data (usage mode 1), \ccs{in practice, a user may aim to defend against a specific subset of sensitive inferences instead of focusing on all possible inferences.} \ccs{For instance, Blowfish privacy~\cite{he2014blowfish} has recently been deployed by the US Census for protecting specific sensitive information.} Therefore, we further consider a scenario where the user aims to defend against a \emph{specific subset of sensitive inferences} over the sensor data (usage mode 2). Under this threat model, there are novel opportunities to strategically add noise and enhance utility. We thus relax the definition of LDP to provide provable privacy guarantees for a specific subset of sensitive inferences but significantly improve the utility performance.}

\unknown{Pufferfish ~\cite{kifer2014pufferfish} and Blowfish~\cite{he2014blowfish} privacy frameworks have similarly motivated the threat model of protecting against a specific set of sensitive computations over the data.}  \ccs{Inspired by these frameworks and Definition~\ref{dpdef}, we define our relaxed neighboring databases as follows.} \ccsnew{The advantage of our formulation is that we provide a computationally efficient mechanism to achieve this rigorous privacy guarantee while the same is not true for Pufferfish privacy.}
\begin{mydef}
\label{neigh_data}
\revise{Let us represent a set of sensitive inferences as ${S}(\cdot)$ \SP{which map the temporal sensor data to their corresponding inference results (i.e., labels), which is the information we
want to protect}. Let us represent an orthogonal function of $S(\cdot)$ as ${K}(\cdot)$. Thus, $\int_{x} {K}(x)\times S(x) dx=0$, where ${K}(\cdot)\in {\bm{{S}}}^{\perp}$ and ${\bm{{S}}}^{\perp}$ is the orthogonal component of ${S}(\cdot)$ consisting of all the functions that are orthogonal to ${S}(\cdot)$. \needcheck{Two databases $D_1, D_2$ are relaxed neighboring databases, iff. ${S}(D_1)\neq {S}(D_2)$, \blue{${K}(D_1)={K}(D_2)$ for any function ${K}(\cdot)$}}.}
\end{mydef}
	
	\begin{figure}[!t]
	\centering
	\includegraphics[width=2.7in,height=0.7in]{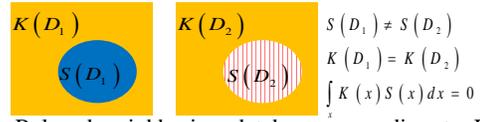}
	\DeclareGraphicsExtensions.
	\vspace{-1em}
	\caption{\unknown{Relaxed neighboring databases according to Definition~\ref{neigh_data}, which have different sensitive inference values whilst the same values for the orthogonal inferences.}}
	\label{nei}
	\vspace{-1em}
	\end{figure}

\revise{\indent We illustrate the relaxed neighboring databases above in Figure~\ref{nei}. For concrete interpretation of the definition, we refer back to Case Study 3 in Section~\ref{sec:intro}, where the useful inference is \emph{speech-to-text translation} and the sensitive inference is \emph{speaker identity recognition}. Intuitively, both the useful and sensitive inferences rely on a limited number of features extracted from the sensor data. The commonly used features for {speech-to-text translation} are {Mel-frequency Cepstral Coefficients (MFCC)}, {spectrogram} and {N-gram}~\cite{lei2013accurate,michalevsky2014gyrophone,zhou2013ibm}, while the popular features for {speaker identity recognition} are {MFCC}, {spectrogram} and the {voice biometrics} of the user (such as her pitch, accent, etc.)~\cite{liu2012cloud,nirjon2013auditeur}. According to Definition~\ref{neigh_data}, the relaxed neighboring databases there refer to two sensor data streams that correspond to different speakers (i.e., different values of MFCC, spectrogram and the voice biometrics) and have the same values of N-gram (the orthogonal component). {Therefore, the relaxed neighboring databases in Definition~\ref{neigh_data} focuses on protecting the privacy of sensitive inferences only (whose values differ in the two databases), and not the component of the data that is orthogonal to sensitive inferences (whose values remain the same in the two databases).}}

\begin{figure}[!t]
	\centering
	\includegraphics[width=2.3in,height=1.1in]{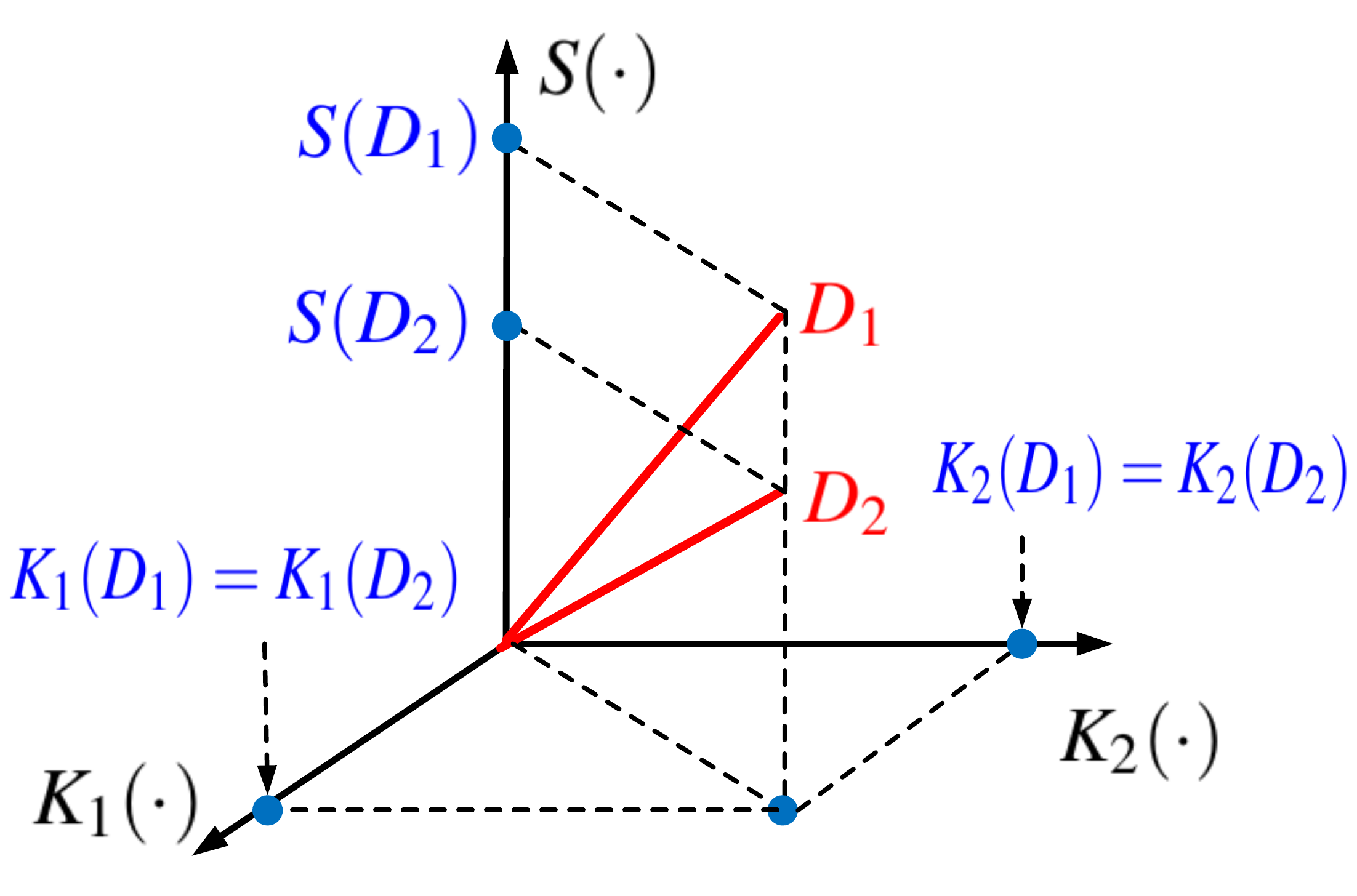}
	\DeclareGraphicsExtensions.
	\caption{\secondrevise{Illustration of the two relaxed neighboring databases $D_1,D_2$ in Definition~\ref{neigh_data}}. $K_1(\cdot), K_2(\cdot)$ are orthogonal functions of the sensitive inference function $S(\cdot)$.}
	\label{illu}
	\vspace{-1em}
	\end{figure}

\noindent\unknown{{\bf{Remark on orthogonal functions:}}} 
\secondrevise{We further illustrate the relaxed neighboring databases in a three-dimensional space (our definition can be generally applied to any space) in Figure~\ref{illu}, where there are two functions $K_1(\cdot), K_2(\cdot)$ that are orthogonal with the sensitive inference function $S(\cdot)$, i.e., $\int_{x} {K}_1(x)\times S(x) dx=0, \int_{x} {K}_2(x)\times S(x) dx=0$. According to Definition~\ref{neigh_data}, we have the following properties for the two relaxed neighboring databases $D_1,D_2$: 1) they have the same projection on the direction corresponding to the orthogonal functions, i.e., $K_1(D_1)=K_1(D_2), K_2(D_1)=K_2(D_2)$; 2) they have different projection on the direction corresponding to the sensitive inference function, i.e., $S(D_1)\neq S(D_2)$.}


\revise{Inspired by the relaxed neighboring databases in Definition~\ref{neigh_data}, we now formulate a relaxed variant of local differential privacy (RLDP) which is also an instantiation of the Pufferfish and Blowfish privacy frameworks. In fact, we can observe that our set of sensitive inferences computed over sensor data is the set of potential secrets defined in Pufferfish privacy (Definition 3.4 in \cite{kifer2014pufferfish}) and Blowfish privacy (Definition 4.2 in \cite{he2014blowfish})}.
\begin{mydef}\label{IPdef}{\emph{($\epsilon$-\usenix{RLDP})}}
\needcheck{For a set of sensitive inference functions} ${S}({\cdot})$ and a privacy budget $\epsilon$, a randomized algorithm ${A}(\cdot)$ satisfies $\epsilon$-\usenix{RLDP} for protecting ${S}({\cdot})$, if for any two relaxed neighboring databases $D_1, D_2$ defined in Definition~\ref{neigh_data} and any output set $o$,
\begin{equation}\label{privacy_def}
\max_{D_1, D_2,o} \frac{\mathbb{P}({A}({D}_1)=o)}{\mathbb{P}({A}({D}_2)=o)}\le \exp(\epsilon)
\end{equation}
Smaller values of $\epsilon$ correspond to higher privacy level.
\end{mydef}
\thirdrevise{To achieve $\epsilon$-\usenix{RLDP} for protecting sensitive information, it is required that the shared data $o={A}({D_1})={A}({D_2})$ conditioned on any two different sensitive information ${S}(D_1), {S}(D_2)$ (where ${S}(D_1)\neq {S}(D_2)$ in Definition~\ref{neigh_data}), is statistically indistinguishable from each other.} This in turn guarantees that an adversary gains negligible information about the true sensitive information upon observing the shared data. We re-iterate that unlike traditional \usenix{LDP} frameworks, its relaxed variant focuses only on the privacy of a specific subset of sensitive inferences. A perturbation mechanism developed to satisfy \usenix{RLDP} guarantees would thus result in better utility performance than the \usenix{LDP} mechanisms.
\\ \indent\ccsnew{By extending the previous results on composition properties for \usenix{LDP} in \cite{duchi2013local}, our RLDP also composes securely, i.e., retains privacy guarantees even in the presence of multiple releases. Furthermore, our RLDP is also immune to post processing. Therefore, a data analyst, cannot increase privacy loss of the data by computing a function of the output of a RLDP-private algorithm.} The detailed proofs for these privacy properties are deferred to the appendix to improve readability.
\begin{mythe}{\bf{(Sequential Composition Theorem)}}\label{seq}
\needcheck{Let randomized algorithm ${A}_t(D)$ ($t=1,2,\cdots$) each provide $\epsilon_t$-\usenix{RLDP} under the sensitive inference functions $S(\cdot)$. The sequence of these algorithms ${A}_t({D})$ provides $\sum\limits_{t} \epsilon_t$-\usenix{RLDP}.}
\end{mythe}
\begin{mythe}{\bf{(Parallel Composition Theorem)}}\label{paral}
\needcheck{Let randomized algorithms ${A}_t(D_t)$ ($t=1,2,\cdots$) provide $\epsilon_t$-\usenix{RLDP} under the sensitive inference functions $S(\cdot)$ and $D_t$ be arbitrary disjoint data set. The sequence of these randomized algorithm ${A}_t(D_t)$ provides $\max\limits_t \epsilon_t$-\usenix{RLDP}.}
\end{mythe}
\begin{mythe}{\bf{\ccs{(Post-processing Invariant Property)}}}\label{ppip}
\ccs{Let randomized algorithm ${A}(D)$ provide $\epsilon$-\usenix{RLDP} under the sensitive inference functions $S(\cdot)$. Let $f$ be an arbitrary mapping computed over $A(D)$. Then, $f(A(D))$ provides $\epsilon$-\usenix{RLDP}.}
\end{mythe}
\ndss{\quad It is worth noting that our proposed RLDP is of independent interest even outside of its application to mobile systems.}	

\section{Perturbation Mechanisms}
\revise{In this section, we design effective perturbation mechanisms to achieve $\epsilon$-LDP and $\epsilon$-RLDP (corresponding to the two usage modes in Section~\ref{sec_threat}) for protecting mobile sensor data in a single-user setting. \emph{\ndss{Our key insight is to exploit the structural characteristics of mobile sensor data to enhance privacy-utility tradeoffs.}} Before proposing our privacy mechanisms, we first introduce a baseline approach which directly generalizes the traditional DP perturbation mechanisms to achieve $\epsilon$-LDP.}
\subsection{\usenix{Baseline Approach}}\label{sec_baseline}
\revise{To achieve $\epsilon$-\usenix{DP}, the Laplace Perturbation Mechanism (LPM)~\cite{dwork:Springer06} applies noise drawn from a suitable Laplace distribution to perturb the query results. More formally, for a query function $Q(\cdot)$, LPM computes and outputs $A(D)=Q(D)+Lap(\lambda)$, where $\lambda=\Delta Q/\epsilon$ is the parameter of the Laplacian noise and $\Delta Q=\max_{D_1,D_2}\|Q(D_1)-Q(D_2)\|_1$ is the global sensitivity of $Q(\cdot)$~\cite{dwork:Springer06}.} 

\revise{When applying LDP (recall Definition~\ref{dpdef}) on mobile sensor data ${\bm{x}}_t$, the neighboring databases ${\bm{x}}_{t1}, {\bm{x}}_{t2}$ may differ in all their possible tuples (i.e., all sensor recordings across all timestamps within the same window), while the neighboring databases in traditional DP frameworks only differ in one tuple. According to the \emph{composition theorem} of DP~\cite{mcsherry2009privacy}, we propose a baseline approach to achieve $\epsilon$-LDP, which inserts a Laplacian noise to each temporal sensor data point with the same parameter of \usenix{$\lambda=\dim({\bm{x}}_t)\Delta Q/\epsilon$}~\footnote{\usenix{Note that the sensitivity $\Delta Q$ is estimated from our collected data serving as a local sensitivity.}}, where $\dim({\bm{x}}_t)$ is the dimension of each segmented sensor data ${\bm{x}}_t$ \secondrevise{(we refer interested readers to  \cite{chen2014correlated} for similar perturbation mechanisms). The baseline approach thus achieves $\epsilon$-LDP guarantees.}}
\subsection{\usenix{DEEProtect Perturbation Mechanisms for Enhanced Privacy-Utility Tradeoffs}}
\revise{Note that the baseline approach introduces noise that is linear in the number of temporal sensor recordings within the time window. This increases the magnitude of noise that needs to be added and thus degrades the usability of the data. Therefore, a better alternative to protect sensor data is to build a compact, privacy-preserving synopsis from the data by exploiting its structural characteristics.}

\revise{Data minimization~\cite{house2012consumer,pfitzmann2010terminology,cavoukian2009privacy} is a fundamental legal instrument that protects privacy by limiting the collection of personal data to the minimum extent necessary for attaining legitimate goals. In our work, we enforce the principle of data minimization and retain only the minimum number of features necessary to enable the authorized inferences. \ccsnew{Specifically, we propose a novel supervised autoencoder model to extract authorized features that are only relevant to the useful inferences. 
 These extracted features may still be correlated with the sensitive inferences; thus we incorporate LDP/RLDP guarantees to obfuscate these features for protecting against sensitive inferences.}} \revise{In this section, we propose DEEProtect (recall Figure~\ref{fig:pipeline}) which consists of two key steps of 1) first extracting features based on the data minimization principle and 2) then perturbing these features to achieve both conventional and relaxed notions of LDP guarantees.}
\subsubsection{\ccs{Autoencoder} based Data Minimization}\label{sec_feature}
{Mobile sensor data is usually high-dimensional in nature, which typically exhibits both structure and redundancy, allowing minimization~\cite{li2013compressed,basch1999data}. This lays the foundation for the first technique in our DEEProtect system: \ccs{autoencoder} based data minimization.}

\noindent{{\bf{Existing Autoencoder Models:}} \ccs{Autoencoders \cite{hinton2006reducing,lee2008sparse} automatically extract features in an unsupervised manner by minimizing the reconstruction error between the input and its reconstructed output. Multi-layer autoencoders learn transformations from the input data to the output representations, which is more powerful for feature extraction than hand-crafted shallow models.} \secondrevise{A single-layer autoencoder is shown in Figure~\ref{auto}. The encoder function $E_{nc}(\cdot)$ maps the input data ${\bm{x}}_t$ to the hidden units (features) according to ${\bm{f}}_t=E_{nc}({\bm{x}}_t)=g_{enc}({\bm{W}}{\bm{x}}_t+{\bm{b}}_e)$, where $g(\cdot)$ is typically a sigmoid function, $\bm{W}$ is a weight matrix and ${\bm{b}}_e$ is a bias vector. The decoder function $D_{ec}(\cdot)$ maps these features back to the original input space according to
$\widetilde{\bm{x}}_t=D_{ec}({\bm{f}}_t)=g_{dec}({\bm{W}}^\prime{\bm{f}}_t+{\bm{b}}_d)$, where $g_{dec}(\cdot)$ is usually the same form as that in the encoder, ${\bm{W}}^\prime$ is a weight matrix  and ${\bm{b}}_d$ a bias vector. Considering the inherent structure of the mobile sensor data~\cite{li2013compressed,basch1999data}, we aim to learn an effective feature space on which the mobile sensor data would have a succinct representation, and the corresponding objective function is as follows.}
	\begin{figure}[!t]
	\centering
	\includegraphics[width=2.8in,height=0.8in]{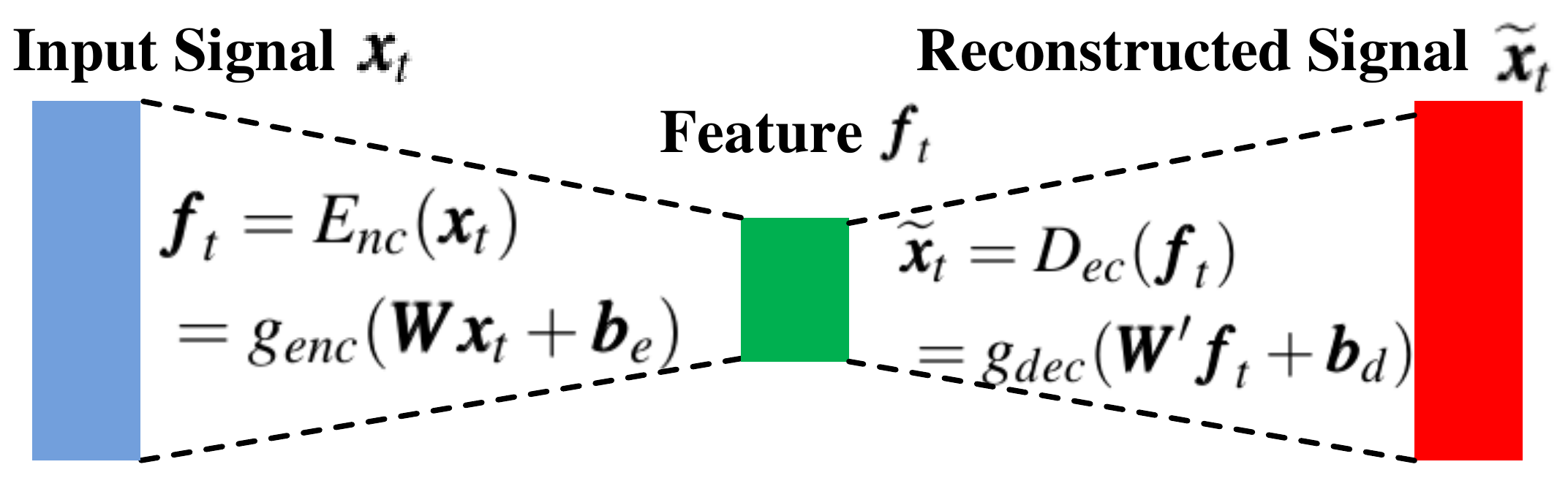}
	\DeclareGraphicsExtensions.
	\caption{\secondrevise{Illustration of Autoencoder Models.}}
	\label{auto}
	\vspace{-1em}
	\end{figure}
\begin{equation}\label{cost}
\begin{aligned}
J_{0}({\bm{W}},{\bm{b}}_e,{\bm{b}}_d,{\bm{W}}^\prime)=\sum\nolimits_{t=1}^{T/N_w} Re\_error({\bm{x}}_t,D_{ec}(E_{nc}({\bm{x}}_t)))
\end{aligned}
\end{equation}
\secondrevise{where $Re\_error({\bm{x}}_t,D_{ec}(E_{nc}({\bm{x}}_t)))$ is the reconstruction error between the input data ${\bm{x}}_t$ and its reconstructed output $D_{ec}(E_{nc}({\bm{x}}_t))$. Existing autoencoders thus aim to minimize $J_{0}({\bm{W}},{\bm{b}}_e,{\bm{b}}_d,{\bm{W}}^\prime)$ with respect to ${\bm{W}},{\bm{b}}_e,{\bm{b}}_d,{\bm{W}}^\prime$~\cite{hinton2006reducing,lee2008sparse}.}

\noindent{\bf{Our New Approach for Data Minimization:}} Using a nonlinear encoding function an autoencoder can typically extract better features than previous linear transformation methods~\cite{van2009dimensionality}. But, the features learnt are not specific to the useful inferences. \unknown{To handle this, we explicitly modify the autoencoder models in Eq.~\ref{cost} by incorporating the useful inferences and other associated constraints as follows.} \unknown{\emph{To the best of our knowledge, this is the first work to modify the \ccs{autoencoder} models through incorporating the authorized inferences.}} 

\noindent{\bf{Incorporating Useful Inferences:}} The objective for our data minimization mechanism is to maximize utility performance with the minimum amount of information, therefore it is important to combine the useful inferences with autoencoder models {to automatically extract features in a supervised manner}. Specifically, we incorporate the minimization of cost function corresponding to useful inferences to the objective function of existing autoencoders in Eq.~\ref{cost}. 

We analyze the cost function for each inference using {Case Study 1} in Section~\ref{sec:intro}, where behavior-based authentication was considered as a useful inference and activity mode detection was deemed sensitive. \emph{\ndss{We emphasize that our autoencoder model only requires knowledge of the useful and not the sensitive inferences -- a requirement exactly similar to traditional DP.}}
\ndss{The useful inferences can be mathematically transformed into classification problems, which can be addressed by machine learning techniques}. For instance, by leveraging the popular \emph{ridge regression} technique~\cite{hoerl1970ridge}, we can learn an optimal classifier as follows.
\begin{equation}\label{krr_obj}
\begin{aligned}
{\bm{c}}^{*{U}}=\arg\min_{{\bm{c}}}\mathbb{C}^{{U}}
=\arg\min_{{\bm{c}}}\beta\|{\bm{c}}\|^2+\sum\nolimits_{t=1}^{T/N_w}({\bm{c}}^T{\bm{x}}_t-y_t^{{U}})^2
\end{aligned}
\end{equation}
\noindent\ndss{where $\mathbb{C}^{*{U}}$ represents the cost function for the useful inference. For behavior-based authentication ({useful inference} in Case Study 1), the label $y_t^{{U}} \in \{1, -1\}$ where $1$ represents the legitimate user and $-1$ represents the adversary\footnote{Note that such labels are only needed for the training process and are not required when a normal user utilizes our DEEProtect system.}. The optimal classifier ${\bm{c}}^{*{U}}$ learned in Eq.~\ref{krr_obj} can be utilized to label the newly-coming mobile sensor data for behavior-based authentication.} \needcheck{The cost function in Eq.~\ref{krr_obj} has been popularly used in machine learning community whilst other general models \ccs{in~\cite{tsochantaridis2004support,bertsekas1995neuro,hinton2006reducing}} can also be explored as potential cost function of these inferences.} Note that our analysis are not restricted to these settings and can be utilized for arbitrary inference-based mobile applications.}

\noindent\usenix{{\bf{Incorporating Orthogonality:}} In addition, we also aim to learn {{orthogonal}} features so that we can deal with each feature independently in our feature perturbation mechanism for achieving enhanced utility performance. Therefore, we add another constraint as ${\bm{W}}{\bm{W}}^T={\bm{I}}$ to the objective function to guarantee the \emph{orthogonality} of the features (similar intuition has also been utilized in~\cite{tipping1999probabilistic}).}

\noindent\usenix{\secondrevise{\bf {Our New Autoencoder Model:}} Incorporating useful inferences and orthogonality into Eq.~\ref{cost} and combining with Eq.~\ref{krr_obj}, we construct the objective function for our data minimization method as
\begin{equation}
\label{final_obj}
\begin{aligned}
&\min J({\bm{W}},{\bm{b}}_e,{\bm{b}}_d,{\bm{W}}^\prime,{\bm{c}})=\min J_{0}({\bm{W}},{\bm{b}}_e,{\bm{b}}_d,{\bm{W}}^\prime)\\
&~~~~~+\mu\beta\|{\bm{c}}\|^2+\mu\sum\nolimits_{t=1}^{T/N_w}({\bm{c}}^T{(D_{ec}(E_{nc}({\bm{x}}_t))}-y_t^{{U}})^2\\
&s.t.~~~ {\bm{W}}{\bm{W}}^T={\bm{I}}
\end{aligned}
\end{equation}
where $\mu$ controls the trade-off between the reconstruction loss and the  utility penalty, and ${\bm{W}}{\bm{W}}^T={\bm{I}}$ represents the orthogonality constraint. \unknown{Note that the last two terms in Eq.~\ref{final_obj} correspond to the cost function in ridge regression in Eq.~\ref{krr_obj}.} Although we only consider the cost function of ridge regression to optimize the data minimization process, our algorithm can be generalized to multiple machine learning techniques such as support vector machine~\cite{suykens1999least}, naive Bayesian~\cite{rish2001empirical} and random forests~\cite{breiman2001random} in a straightforward manner.}

\begin{algorithm}[!t]
\renewcommand{\algorithmicrequire}{\textbf{Input:}}
\renewcommand\algorithmicensure {\textbf{Output:} }
\begin{algorithmic}[1]
\REQUIRE~~{ 
{The sensor data $\{{\bm{x}}_t\}_{t=1}^{T/N_w}$, the learning rate $\alpha$;}}\\
\ENSURE {
{Weight matrix $\bm{W}, {\bm{W}}^\prime$ and bias ${\bm{b}}_e, {\bm{b}}_d$};\\    
       1. Initialize ${\bm{W}},{\bm{b}}_e, {\bm{b}}_d, {\bm{c}}$ randomly and set ${\bm{W}}^\prime={\bm{W}}^T$;\\
       2. {\bf{For each iteration}} $=1,2,3\cdots$ {\bf{do}}\\
       3.~~~~~~ Set ${{\Delta}} {\bm{W}}=0, {{\Delta}} {\bm{b}}_e=0, {\bm{\Delta}} {\bm{b}}_d=0, {{\Delta}} {\bm{c}}=0$;\\
       4.~~~~~~ Compute the partial derivatives:
       \begin{equation}
       \begin{footnotesize}
       {
       \begin{aligned}
       &\Delta{\bm{W}}=\frac{\partial J({\bm{W}},{\bm{b}}_e,{\bm{b}}_d,{\bm{W}}^\prime,{\bm{c}})}{\partial {\bm{W}}},~~~\Delta{\bm{b}}_e=\frac{\partial J({\bm{W}},{\bm{b}}_e,{\bm{b}}_d,{\bm{W}}^\prime,{\bm{c}})}{\partial {\bm{b}}_e}\\
       &\Delta{\bm{b}}_d=\frac{\partial J({\bm{W}},{\bm{b}}_e,{\bm{b}}_d,{\bm{W}}^\prime,{\bm{c}})}{\partial {\bm{b}}_d},~~~\Delta{\bm{c}}=\frac{\partial J({\bm{W}},{\bm{b}}_e,{\bm{b}}_d,{\bm{W}}^\prime,{\bm{c}})}{\partial {\bm{c}}}
       \end{aligned} }
       \end{footnotesize}
       \end{equation} 
       5.~~~~~~ Update ${\bm{W}},{\bm{b}}_e,{\bm{b}}_d,{\bm{c}}$ by gradient descent as
       \begin{equation}
       \begin{footnotesize}
       {
       \begin{aligned}
      {\bm{W}}&:={\bm{W}}-\alpha\Delta{\bm{W}}, &{\bm{b}}_e&:={\bm{b}}_e-\alpha\Delta{\bm{b}}_e\\ {\bm{b}}_d&:={\bm{b}}_d-\alpha\Delta{\bm{b}}_d, &~~{\bm{c}}&:={\bm{c}}-\alpha\Delta{\bm{c}}~~~
       \end{aligned}}
       \end{footnotesize}
       \end{equation}
       6.~~~~~~ Orthogonalize the weight matrix as}
       \begin{equation}
       {
       {\bm{W}}:=({\bm{WW}}^T)^{-\frac{1}{2}}\bm{W}}
       \end{equation}
\caption{
{Autoencoder based Data Minimization}}
\label{alg1}
\vspace{-1em}
\end{algorithmic}
\end{algorithm}

\noindent\usenix{{\bf{Model Learning and Stacking:}} To minimize $J({\bm{W}},{\bm{b}}_e,{\bm{b}}_d,{\bm{W}}^\prime, {\bm{c}})$ in Eq.~\ref{final_obj} with respect to ${\bm{W}},{\bm{b}}_e,{\bm{b}}_d,{\bm{W}}^\prime,{\bm{c}}$, we explore the stochastic gradient descent (SGD) technique, which has been shown to perform fairly well in practice~\cite{bottou1991stochastic}. Furthermore, we explore multiple hidden layers to stack multiple model units, in order to generate even more compact and higher-level semantic features resulting in better data minimization. \revise{To improve readability, we defer all the details about model learning and stacking into the appendix, based on which we summarize our \ccs{autoencoder} based data minimization mechanism in Algorithm~\ref{alg1}.}}
\subsubsection{\revise{Perturbation Mechanism for Usage Mode 1 (Achieving $\epsilon$-LDP)}}\label{sec:usage1}
\revise{LDP considers the worst-case adversary which can rigorously protect \emph{against all possible inferences} computed over the data (recall Definition~\ref{dpdef}). In the absence of \ndss{system-specified} set of sensitive inferences, or otherwise if the user chooses to operate under the LDP guarantees (usage mode 1 in DEEProtect), we develop our perturbation mechanism through perturbing the features learnt from the \ccs{autoencoder} based data minimization step in Section~\ref{sec_feature}. Formally, to achieve $\epsilon$-LDP, DEEProtect under usage mode 1 inserts an Laplacian noise with parameter \usenix{$\lambda=\sqrt{\dim({\bm{f}}_t)}\dim({\bm{x}}_t)\Delta Q /\epsilon$} to each previously learned feature, where $\dim({\bm{x}}_t), \dim({\bm{f}}_t)$ represent the dimension of the segmented sensor data ${\bm{x}}_t$ and the features ${\bm{f}}_t$ extracted from ${\bm{x}}_t$, respectively.\footnote{\unknown{Although the noise parameter is higher than the baseline approach with a factor of $\sqrt{\dim({\bm{f}}_t)}$, the number of features, i.e., $\dim({\bm{f}}_t)$, needed to be perturbed is much smaller than the number of raw sampling points, i.e., $\dim({\bm{x}}_t)$, in the baseline approach.}} DEEProtect mechanism under usage mode 1 is summarized in Algorithm~\ref{alg3}, which satisfies rigorous $\epsilon$-LDP as will be discussed in Theorem~\ref{proofprivacy}.}

\begin{algorithm}[!t]
\renewcommand{\algorithmicrequire}{\textbf{Input:}}
\renewcommand\algorithmicensure {\textbf{Output:} }
\begin{algorithmic}[1]
\REQUIRE {{Original Sensor Data $\{{\bm{x}}_t\}_{t=1}^{T/N_w}$; Useful Inference Sets ${U}(\cdot)$; Privacy Budget $\epsilon$;}}\\ 
\ENSURE  \revise{{Perturbed Sensor Data $\{{\bm{x}}_t^\prime\}_{t=1}^{T/N_w}$;}\\
       {\bf{For each}} $t=1,2,\cdots,T/N_w$\\
       1. extract features ${\bm{f}}_t$ from ${\bm{x}}_t$ by data minimization mechanism in Algorithm~\ref{alg1};\\
       2. obtain perturbed features ${\bm{f}}_t^\prime$ by inserting noise
       of $Lap(\sqrt{\dim({\bm{f}}_t)}\dim({\bm{x}}_t)\Delta Q /\epsilon)$;\\
       3. reconstruct perturbed sensor data ${\bm{x}}_t^\prime=D_{ec}({\bm{f}}_t^\prime)$;}
\caption{{{DEEProtect under Usage Mode 1.}}}
\label{alg3}
\end{algorithmic}
\end{algorithm}

\revise{Our mechanism is different from the baseline approach in Section~\ref{sec_baseline} because we add Laplacian noise after the application of the \ccs{autoencoder} based data minimization step, while the baseline approach directly adds Laplacian noise to the raw sensor data without the \ccs{autoencoder} mechanism. After perturbing these features, we reconstruct the perturbed sensor data according to the decoder function $D_{ec}(\cdot)$ in autoencoder (recall Eq.~\ref{cost}). We will show that our mechanism significantly outperforms the baseline approach (in Sections~\ref{sec_thory}, \ref{sec_evaluation}). {Note that our privacy objective is also different from that in~\cite{shokri2015privacy,abadi2016deep} since they aim to protect each user's training data in the \ccs{autoencoder} training stage under the multi-user settings while in contrast we aim to protect the privacy of mobile sensor data stream in single-user settings.}}


\subsubsection{\usenix{Perturbation Mechanism for Usage Mode 2 (Achieving $\epsilon$-RLDP)}}\label{sec:usage2}
{If the \ndss{system configuration} has specified a subset of sensitive inferences (usage mode 2 in DEEProtect), we develop our perturbation mechanism (summarized in Algorithm~\ref{alg2}) by first computing the relaxed sensitivity $\Delta Q_{\mathrm{relax}}$, and then inserting an Laplacian noise with parameter $\lambda_{\mathrm{relax}}=\sqrt{\dim({\bm{f}}_t)}\dim({\bm{x}}_t)\Delta Q_{\mathrm{relax}} /\epsilon$ to each previously learned feature, to satisfy $\epsilon$-RLDP. The detailed process to compute the relaxed sensitivity is as follows.}

\noindent\revise{{\bf{Computing Relaxed Sensitivity:}}} 
Similar to \emph{sensitivity} in DP~\cite{dwork:Springer06}, our \emph{relaxed sensitivity} can be computed as
\begin{equation}\label{relaxed_sensitivity0}
\Delta Q_{\mathrm{relax}}={\max_{{\bm{x}}_{t1}, {\bm{x}}_{t2}}\|Q({\bm{x}}_{t1})-Q({\bm{x}}_{t2})\|_1}/{{\dim({\bm{x}}_t)}}
\end{equation}
where ${{\bm{x}}_{t1}, {\bm{x}}_{t2}}$ are relaxed neighboring databases in Definition~\ref{neigh_data}, and the denominator is set to make \emph{relaxed sensitivity} comparable with traditional \emph{sensitivity} computed over neighboring databases that differ in only one tuple.

\begin{algorithm}[!t]
\renewcommand{\algorithmicrequire}{\textbf{Input:}}
\renewcommand\algorithmicensure {\textbf{Output:} }
\begin{algorithmic}[1]
\REQUIRE {{Original Sensor Data $\{{\bm{x}}_t\}_{t=1}^{T/N_w}$; Useful and Sensitive Inference Sets ${U}(\cdot), {S}(\cdot)$; Privacy Budget $\epsilon$;}}\\ 
\ENSURE  \revise{{Perturbed Sensor Data $\{{\bm{x}}_t^\prime\}_{t=1}^{T/N_w}$;}\\
       {\bf{For each}} $t=1,2,\cdots,T/N_w$\\
       1. extract features ${\bm{f}}_t$ from ${\bm{x}}_t$ by data minimization mechanism in Algorithm~\ref{alg1};\\
       2. compute relaxed sensitivity $\Delta Q_{\mathrm{relax}}$ in Eq.~\ref{relaxed_sensitivity0};\\
       3. obtain perturbed features ${\bm{f}}_t^\prime$ by inserting noise of $Lap(\sqrt{\dim({\bm{f}}_t)}\dim({\bm{x}}_t)\Delta Q_{\mathrm{relax}} /\epsilon)$;\\
       4. reconstruct perturbed sensor data ${\bm{x}}_t^\prime=D_{ec}({\bm{f}}_t^\prime)$;}
\caption{{{DEEProtect under Usage Mode 2.}}}
\label{alg2}
\end{algorithmic}
\end{algorithm}

\secondrevise{The sensitive inferences can be mathematically transformed into a classification problem, which can be addressed by machine learning techniques. \ccsnew{\emph{Specifically, we can also infer the sensitive inference classifier as the ridge regression classifier in Eq.~\ref{krr_obj}, by utilizing the labeled training data corresponding to the sensitive inference function.}} \ndss{Let us take the {activity mode detection} ({sensitive inference} in Case Study 1) as an example. The corresponding labels $y_t^{{S}}$ for the mobile sensor data ${{\bm{x}}_t}$ satisfy $y_t^{{S}} \in \{1, 2, 3\}$ where $1,\;2$ and $3$ represent \emph{walking}, \emph{standing still} and \emph{moving up or down the stairs}, respectively. The ridge regression classifier  ${\bm{c}}^{*{S}}$ for the sensitive inference is thus a linear function computed over the segmented sensor data ${\bm{x}}_t$, i.e., ${S}({\bm{x}}_t)={{\bm{x}}_t}^T\cdot{\bm{c}}^{*{S}}$.} We explain the computation of relaxed sensitivity using this formulation, though our analysis can be generally applied to non-linear situations using kernel-based techniques~\cite{cristianini2000introduction}. We apply the \emph{Gram-Schmidt orthogonalization} technique \cite{bjorck1967solving} to a matrix $[{\bm{c}}^{*{S}};{\bm{I}}]$ (where $\bm{I}$ is an \emph{identity} matrix and thus $[{\bm{c}}^{*{S}};{\bm{I}}]$ is full-rank), in order to obtain orthogonal vectors of ${\bm{c}}^{*{S}}$ as ${{\bm{c}}^{*{S}}_{\perp1}},{{\bm{c}}^{*{S}}_{\perp2}},\cdots$. Based on that, we form an orthogonal matrix ${\bm{S}}=[{{\bm{c}}^{*{S}}};{{\bm{c}}^{*{S}}_{\perp1}};{{\bm{c}}^{*{S}}_{\perp2}};\cdots]$. Any function ${K}(\cdot)$ that is orthogonal of the sensitive inference function ${S}(\cdot)$ (recall Definition~\ref{neigh_data}) can be represented by a linear combination of ${{\bm{c}}^{*{S}}_{\perp1}},{{\bm{c}}^{*{S}}_{\perp2}},\cdots$.}

\secondrevise{For the two relaxed neighboring databases ${\bm{x}}_{t1}, {\bm{x}}_{t2}$, we have ${\bm{x}}_{t1}^T\cdot{{\bm{c}}^{*{S}}_{\perp1}}={\bm{x}}_{t2}^T\cdot{{\bm{c}}^{*{S}}_{\perp1}}, {\bm{x}}_{t1}^T\cdot{{\bm{c}}^{*{S}}_{\perp2}}={\bm{x}}_{t2}^T\cdot{{\bm{c}}^{*{S}}_{\perp2}}, \cdots$, according to Definition~\ref{neigh_data}. Therefore, we know that ${\bm{S}}\cdot({\bm{x}}_{t1}-{\bm{x}}_{t2})={\bm{S}}\cdot{\bm{x}}_{t1}-{\bm{S}}\cdot{\bm{x}}_{t2}=[{\bm{x}}_{t1}^T\cdot{\bm{c}}^{*{S}}-{\bm{x}}_{t2}^T\cdot{\bm{c}}^{*{S}}, {\bm{x}}_{t1}^T\cdot{\bm{c}}^{*{S}}_{\perp1} -{\bm{x}}_{t2}^T\cdot{\bm{c}}^{*{S}}_{\perp1},{\bm{x}}_{t1}^T\cdot{\bm{c}}^{*{S}}_{\perp2}-{\bm{x}}_{t2}^T\cdot{\bm{c}}^{*{S}}_{\perp2},\cdots]^T=[\gamma, 0,0,\cdots]^T$, i.e., ${\bm{x}}_{t1}-{\bm{x}}_{t2}={\bm{S}}^{-1}\cdot[\gamma,0,0,\cdots]^T$, and the value of $\gamma$ is restricted by the range of ${\bm{x}}_{t}$. Since we consider the \emph{identity} query (as we publish the sanitized sensor data to mobile apps), we obtain the constraint of $\gamma$ as $\|{\bm{S}}^{-1}\cdot[\gamma, 0,0,\cdots]^T\|_1=\|{\bm{x}}_{t1}-{\bm{x}}_{t2}\|_1\le\dim({\bm{x}}_t)\Delta Q$. Therefore, we can compute the relaxed sensitivity as follows.} 
\begin{equation}\label{relaxed_sensitivity}
\small 
\Delta Q_{\mathrm{relax}}=\max\limits_{{\gamma:~\|{\bm{S}}^{-1}\cdot[\gamma,0,0,\cdots]^T\|_1\le \dim({\bm{x}}_t)\Delta Q}}\frac{\|{\bm{S}}^{-1}\cdot[\gamma,0,0,\cdots]^T\|_1}{\dim({\bm{x}}_t)}
\end{equation}
\revise{It is worth noting that $\Delta Q_{\mathrm{relax}}\le\Delta Q$ due to the constraint of relaxed neighboring databases for achieving the same orthogonal inference values (recall Definition~\ref{neigh_data}). Therefore, comparing to usage mode 1, DEEProtect under usage mode 2 would add less noise to the features extracted in the data minimization step thus achieving better utility.} 

\ndss{Our perturbation mechanism in DEEProtect has broad applicability even outside of its application to mobile systems as considered in this paper.}

\section{Theoretical Analysis}\label{sec_thory}
\ccsnew{Next, we theoretically show that our DEEProtect mechanisms satisfy rigorous privacy guarantees and achieve significant utility advantage over the baseline approach.}
We refer the interested readers to the appendix for proofs of our theorems.
\subsection{Theoretical Privacy Analysis} 
\revise{Our perturbation mechanisms in Algorithms~\ref{alg3},~\ref{alg2} satisfy $\epsilon$-LDP, $\epsilon$-RLDP respectively, according to theorems below.}
\begin{mythe}\label{proofprivacy}
\revise{DEEProtect under usage mode 1 (corresponding to Algorithm~\ref{alg3}) satisfies $\epsilon$-LDP.} 
\end{mythe} 
\begin{mythe}\label{proofprivacy2}
\revise{DEEProtect under usage mode 2 (corresponding to Algorithm~\ref{alg2}) satisfies $\epsilon$-RLDP.} 
\end{mythe} 
\subsection{Theoretical Utility Analysis}
\label{utility_priority}
For a perturbation algorithm $A$, let us denote $Error({A})=\mathbb{E}_{{A}}[\|{A}({D})-{D}\|_1]$ as the expected error in the release of data $D$, where $\mathbb{E}_{{A}}[\cdot]$ is the expectation taken over the randomness of ${A}$. \revise{We quantify the utility advantage of DEEProtect over the baseline approach in the following Theorems~\ref{the_error}, \ref{the_error2} (corresponding to the two usage modes).} 
\begin{mythe}
\label{the_error}
\revise{For DEEProtect under usage mode 1 (corresponding to Algorithm~\ref{alg3}), the expected error $Error({A})$ is lower than that of the baseline approach in Section~\ref{sec_baseline} by a factor of $\frac{\dim({{\bm{x}}_t})}{\dim({{\bm{f}}_t})}$, where ${\bm{f}}_t$ is the feature set extracted from the segmented sensor data ${\bm{x}}_t$ by using our \ccs{autoencoder} based data minimization approach.}
\end{mythe}
\begin{mythe}
\label{the_error2}
\revise{For DEEProtect under usage mode 2 (corresponding to Algorithm~\ref{alg2}), the expected error $Error({A})$ is lower than that of the baseline approach  in Section~\ref{sec_baseline} by a factor of $\frac{\dim({{\bm{x}}_t})}{\dim({{\bm{f}}_t})}\cdot\frac{\Delta Q}{\Delta Q_{\mathrm{relax}}}$, where $\Delta Q, \Delta Q_{\mathrm{relax}}$ are the sensitivity and the relaxed sensitivity corresponding to the query function $Q(\cdot)$, respectively.}
\end{mythe}

\indent{Therefore, we know that the utility advantage of our \ccs{autoencoder} based data minimization in Section~\ref{sec_feature} is $\frac{\dim({{\bm{x}}_t})}{\dim({{\bm{f}}_t})}$ and of our relaxed variant of LDP in Section~\ref{sec:usage2} is $\frac{\Delta Q}{\Delta Q_{\mathrm{relax}}}$.}

\section{Experimental Setup}
\secondrevise{In this section, we describe our methodology for collecting sensor data from mobile devices, and the experimental setup (including system parameters) for evaluation.}
\subsection{Sensor Data Collection}
\label{cases}
\needcheck{In our experimentsm, we collected data using a Google Nexus 5 (2.3GHz, Krait 400 processor, 16GB internal storage, 2GB RAM on Android 4.4) and a Moto360 smartwatch (OMAP 3 processor, 4GB internal storage, 512MB RAM on Android Wear OS). On the smartphone, data was captured from the \emph{accelerometer}, \emph{gyroscope}, \emph{orientation}, \emph{ambient light}, \emph{proximity}, and \emph{microphone} sensors. On the smartwatch, the  \emph{accelerometer} and \emph{gyroscope} sensors were recorded (recall Figure~\ref{twodata}). The sampling rate was fixed at \usenix{$10$ Hz}. $20$ \ccs{participants} ($15$ males and $5$ females) were invited to take our smartphone and smartwatch for two weeks and \SP{use them as they would use their personal devices in their daily lives}\footnote{Note that our DEEProtect is not restricted to the usage of these devices and can be generally applied to many real world scenarios.}. \SP{We followed the ethical norms and procedures of our anonymized organization while conducting the user study}.} 

{To obtain the ground-truth information for performance evaluation, we ask the users to record labels for both the \emph{useful} and \emph{sensitive inferences}. \SP{Note that these ground-truth labels are only needed for the training process, and are not required for a normal user to utilize our DEEProtect system.} \ndsstwo{The labeled training data is grouped under two different categories: identity-recognition data or mode-detection data. The identity-recognition data is used for authentication (useful inference in Case Study 1) and speaker identity recognition (sensitive inference in Case Study 3). The labeled data is generated by associating the identity of a user as label to a mobile device, on first use. The mode-detection data correspond to labeled user activities. The number of tasks each user performs is determined by the specific experiment and the corresponding data segments are then labeled as per the tasks. For activity mode detection (sensitive inference in Case Study 1), a user performs 3 tasks: walking, standing still, moving up/down stairs. For transportation detection (useful inference in Case Study 2), a user performs 3 tasks: walking, taking a train, taking a car. For recognition of users' entered text in the keyboard (sensitive inference in Case Study 2), a user performs 36 tasks corresponding to entering the 10 digits and 26 letters, respectively. For speech-to-text translation (useful inference in Case Study 3), a user performs 36 tasks of enunciating 10 digits and 26 letters, respectively.}}

\SP{{\bf{User Study Limitations:}} While the user study described above enables DEEProtect's validation using real-world data, it does have a few limitations. First, we only collect 20 users' data for 2 weeks. It is possible that a larger amount of data from more users would further enhance the performance of DEEProtect. Second, the ground-truth information acquired from users (only needed in the training process)
may still cause bias in statistics in the utility of DEEProtect. Despite these limitations, the experimental results from the user study which will be presented in Section~\ref{sec_evaluation} verify
 the effectiveness of DEEProtect system.} 
\subsection{System Parameter Setup}\label{para}
{\ndss{In our experiments, we segment the temporal mobile sensor data (discussed above) according to the parameter of time window size $N_w=10$}. For the \ccs{autoencoder} based data minimization step (in Section~\ref{sec_feature}),} we use 10-fold cross validation to generate the training data and testing data, where $9/10$ of our collected data is used as training data and the remaining $1/10$ is used as testing data. We repeated this process for $1000$ iterations and reported the averaged results. In our experiments, we used stacked autoencoders ($\alpha=0.1$ in Algorithm~\ref{alg1}) with two hidden layers\footnote{The two hidden layers are most commonly used in existing \ccs{autoencoder} research.} comprising of 15 and 7 units respectively. We will show that an autoencoder with only two-hidden layers was able to extract better features than the state-of-the-art techniques. The reduced number of layers (and units) in the autoencoder allowed us to train the model using small amount (2 weeks) of labeled data from the user (note that we are protecting the sensitive inferences for a single user). \ndss{Note that the training itself is implemented offline and only the trained models/parameters are transferred to the phone. Thus, a privacy-preserving training process in~\cite{ li2013efficient,de2012preserving,cornelius2008anonysense} is orthogonal to our method.} \secondrevise{We implemented all the three case studies (tradeoff between authentication and activity recognition, tradeoff between transportation detection and text recognition, tradeoff between speech translation and speaker identification) discussed in Section~\ref{sec:intro} on our real-world dataset using the system parameters discussed above.}

\section{Evaluation}\label{sec_evaluation}
\revise{In this section, we experimentally demonstrate the effectiveness of DEEProtect using multiple real-world datasets. We first show the advantage of \ccs{autoencoder}-based data minimization over existing feature extraction approaches with up to 2x improvement. Next, we show the advantage of our end-to-end methods combining both feature extraction and perturbation over the baseline approach with up to 8x improvement.}
\subsection{Evaluation for \ccs{Autoencoder}-Based Data Minimization}
\label{feature_exp}
\usenix{We experimentally evaluate our \ccs{autoencoder} based data minimization mechanism in Section~\ref{sec_feature}, using the dataset \needcheck{collected for Case Study 1 (recall Section~\ref{sec:intro}), where the useful inference is the \emph{behavior-based authentication}. To show the advantage of our method,  we further compare it with the state-of-the-art feature extraction approaches.} The discrete \emph{Fourier} transform (DFT) and discrete cosine transform (DCT) are two basic transformation techniques in the signal processing community, and the Haar basis forms the fundamental wavelet transformation for time-frequency signal analysis. \secondrevise{The principal component analysis (PCA) technique \cite{tipping1999probabilistic} can also be utilized to reveal hidden structure in mobile sensor data.} Furthermore, we also consider blind compressive sensing (BCS) as a typical dictionary learning method for comparison~\cite{gleichman2011blind}.} 

\begin{figure}[!t] \centering
\includegraphics[width=3.2in,height=1.2in]{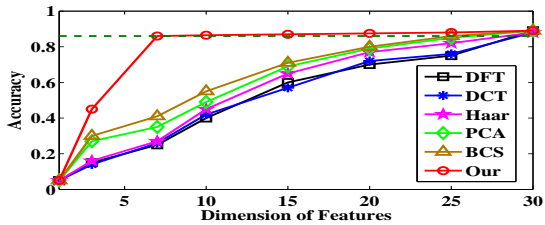}
\vspace{-0.5em}
\caption{\revise{The accuracy for behavior-based authentication in Case Study 1 after data minimization shows the advantage of our method over previous approaches.}}
\label{feature_utility} 
\vspace{-1em}
\end{figure}

\noindent{\bf{Higher Accuracy}}: Figure~\ref{feature_utility} shows the utility-preserving performance under different feature extraction methods. \ccs{Note that $y$-axis is the accuracy for behavior-based authentication in Case Study 1 representing the ratio of correctly authenticated users.} \secondrevise{Note that we use three-dimensional \emph{accelerometer} measurement for behavior-based authentication and set the window size as $N_w=10$ (recall Section~\ref{para}), therefore the dimension of each segmented sensor data is $3\times 10$}. 

From Figure~\ref{feature_utility}, we can see that 1) more features would be beneficial for improving the utility performance since the combination of multiple features would be more accurate and expressive to represent the input data; 2) \ndsstwo{\emph{our method achieves higher accuracy than the state-of-the-art approaches {with up to $2x$ improvement}}. For example, when using $7$ features, our method can provide authentication accuracy of {$86.13\%$} while the accuracy for previous approaches is lower than $41.02\%$, which also validates the effectiveness of the ridge regression classifier explored in our autoencoder-based data minimization model (recall Eq.~\ref{final_obj});} 3) our data minimization mechanism significantly benefits from the automatic learning process \needcheck{which explicitly incorporates the useful inference information into \ccs{autoencoder} models}.
\begin{figure}[!t] \centering
\includegraphics[width=3.2in,height=1.2in]{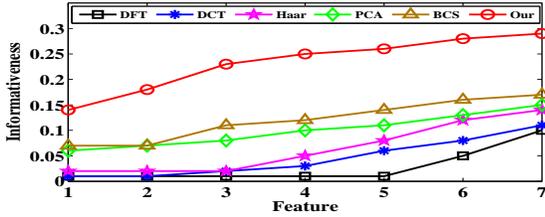}
\vspace{-0.5em}
\caption{\emph{Informativeness}~\cite{frank2013touchalytics} for the features learned from data minimization. Our method produces better features with higher \emph{informativeness} than the state-of-the-art approaches.}
\label{informativeness} 
\vspace{-1em}
\end{figure}

\secondrevise{Based on our analysis above, we constrain the dimension of features extracted by our \ccs{autoencoder} based data minimization method to $7$, whose corresponding accuracy is higher than $97\%$ ($86.13\%/88.39\%$) of the accuracy achieved by using all the $30$ features (with $88.39\%$ accuracy). Therefore, we represent each of the $30$-dimensional input sensor data ${{\bm{x}}_t}$ with a feature set ${\bm{f}}_t$ consisting of only $7$ features.}


\noindent{\bf{Higher Informativeness}}: \usenix{We further leverage \emph{informativeness} in~\cite{frank2013touchalytics} as an important metric to evaluate the information-theoretic relationship between each individual feature and the {useful inference information}.} The \emph{Informativeness} $\mathrm{Info}{(f_i)}$ of the $i$-th feature $f_i$, measures the relative mutual information between the feature and the label of the utility function $y^{{U}}$, and is computed as \usenix{$\mathrm{Info}{(f_i)}=\frac{I(f_i;y^{{U}})}{H(y^{{U}})}=1-\frac{H(y^{{U}}|f_i)}{H(y^{{U}})}$}\footnote{$f_i$ is the random variable taking the $i$-th value in each $\bm{f}_t$.}. Under the setting of {Case Study 1}, $y^{{U}}=1,-1$ represents the legitimate user and the adversary, respectively. $H(y^{{U}})$ is the entropy of the variable $y^{{U}}$ and $I(f_i;y^{{U}})$ is the mutual information between the random variables of $f_i$ and $y^{{U}}$. For each feature $f_i$, this measure of \emph{informativeness} takes a value between $0$ and $1$, where $0$ means that the feature contains no information about the utility label $y^{{U}}$ and $1$ means that the feature can completely determine the utility label $y^{{U}}$. 

\begin{figure}[!t] \centering
\includegraphics[width=3.2in,height=2in]{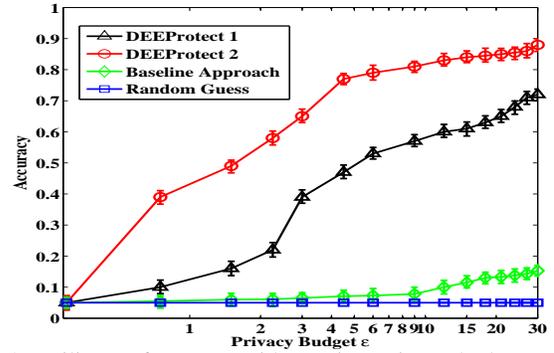}
\vspace{-1em}
\caption{Utility performance with varying privacy budget $\epsilon$ for different methods, \unknown{where \emph{DEEProtect 1} and \emph{DEEProtect 2} represent perturbation mechanism under usage mode 1 (Algorithm~\ref{alg3}) and usage mode 2 (Algorithm~\ref{alg2}), respectively.} A larger privacy budget $\epsilon$ (less perturbation) results in higher accuracy. Furthermore, compared to the baseline approach our perturbation mechanism offers a different tradeoff point which allows 8x improvement in utility. \SP{The range for the 95\% confidence interval of our accuracy results is less than 4.83\%, providing additional validation of our results.}}
\label{acc_epsilon} 
\vspace{-1em}
\end{figure}

\secondrevise{Recall that we set the dimension of features extracted by our method to $7$ (see Figure~\ref{feature_utility}). We further evaluate their corresponding \emph{informativeness} in Figure~\ref{informativeness}.} \emph{\ndss{We can observe that the features learned by our method have much higher informativeness with up to $2x$ improvement over previous works}.} Therefore, our proposed method captures more expressive, higher-quality features than the state-of-the-art approaches.

	\begin{figure*}[!t] \centering
	\subfigure[Utility-Privacy tradeoff for Case Study 1]{
\label{balance1} 
\includegraphics[width=2.4in,height=1.2in]{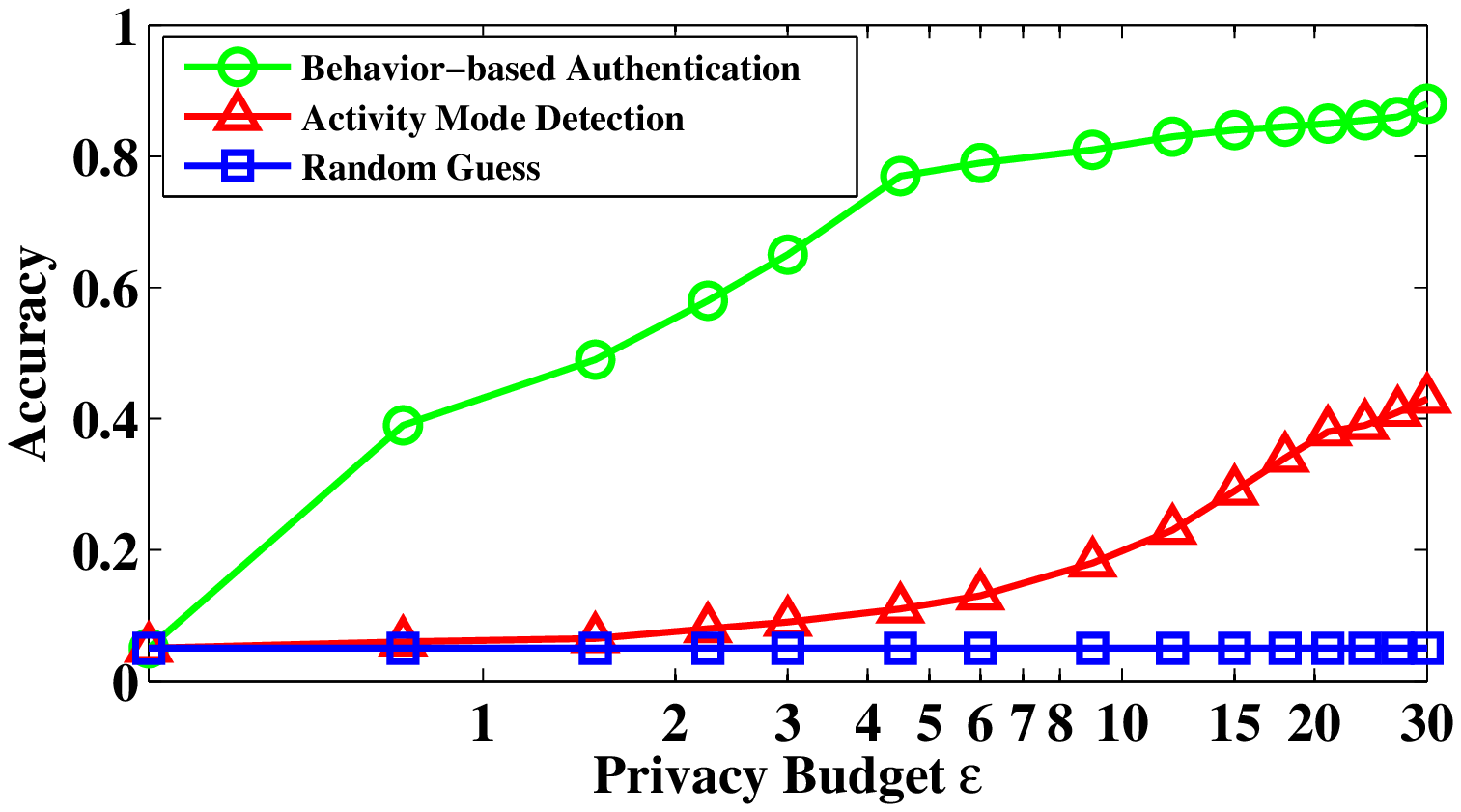}}
\hspace{-0.29in}
\subfigure[Utility-Privacy tradeoff for Case Study 2]{
\label{balance2} 
\includegraphics[width=2.4in,height=1.2in]{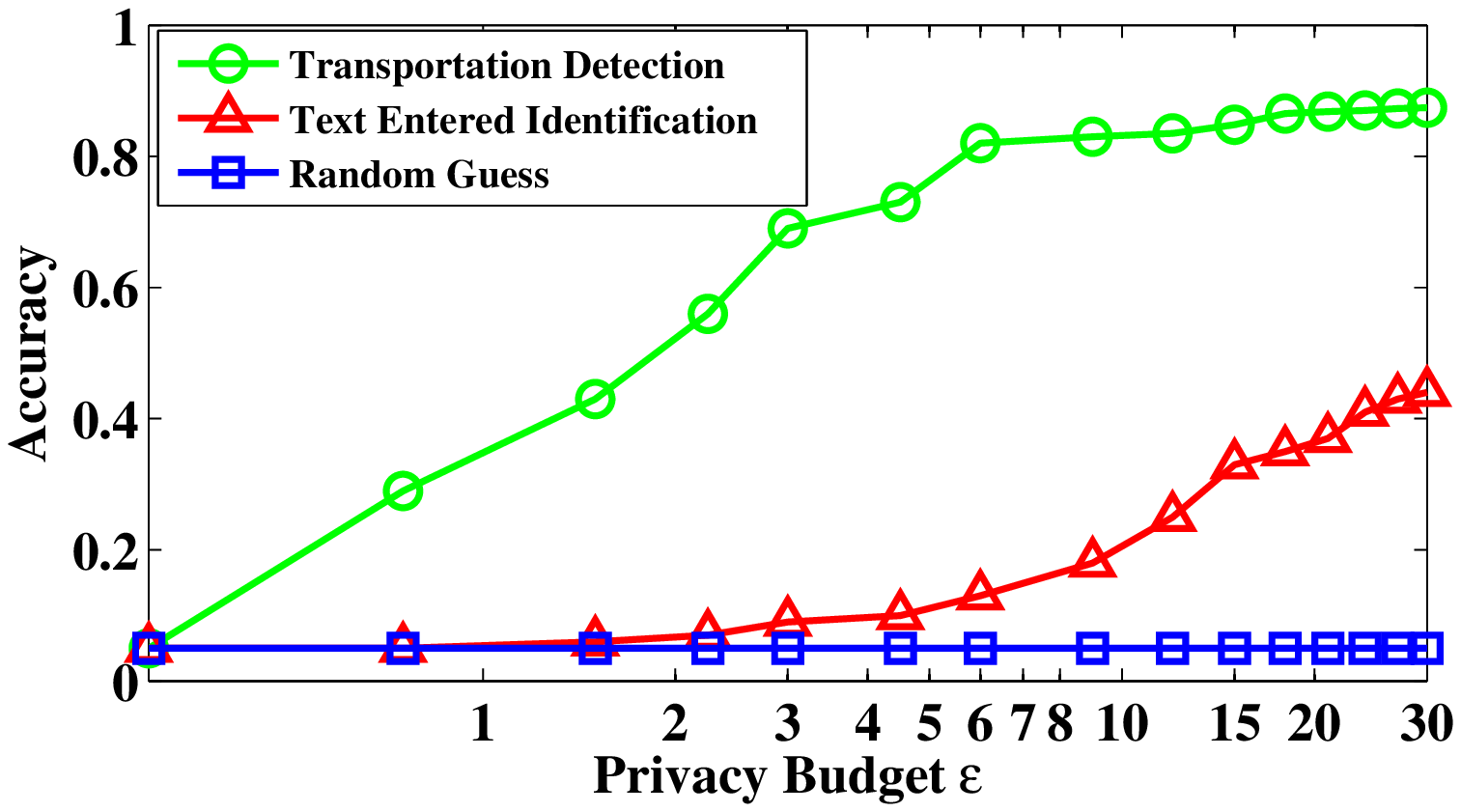}}
\hspace{-0.29in}
\subfigure[Utility-Privacy tradeoff for Case Study 3]{
\label{balance3} 
\includegraphics[width=2.4in,height=1.2in]{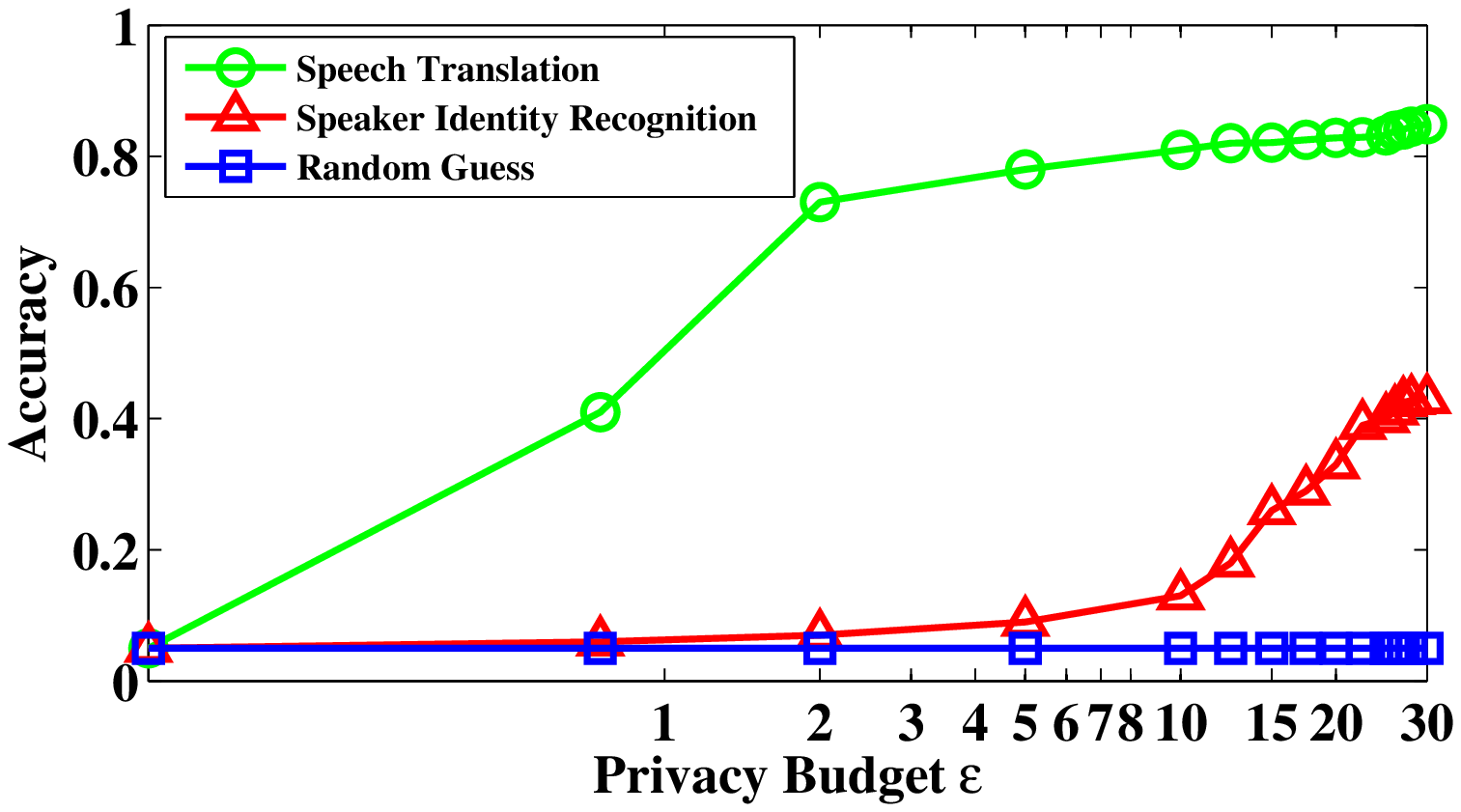}}
\vspace{-1em}
\caption{\unknown{The trade-off between utility and privacy for the three case studies in Section~\ref{sec:intro}. Compared to the useful inferences (shown as green lines), the degradation in accuracy of the {sensitive inference} (shown as red lines) is much faster. \ccs{For $\epsilon=5$, the accuracy for the useful inference is close to the noise free scenario (less than 10\% degradation) while the sensitive inference is close to a random guess}.}}
\label{balance123}
\vspace{-1em}
\end{figure*}

\subsection{Evaluation for DEEProtect End-to-end Mechanisms}
\label{per_exp}
\revise{To demonstrate the effectiveness of our end-to-end perturbation mechanisms combining both feature extraction and perturbation, we again consider {Case Study 1} in Section~\ref{sec:intro} where the \emph{behavior-based authentication} is the useful inference and the \emph{activity mode detection} is the sensitive inference}. We first compare the baseline method with our mechanism under usage mode 1 as shown in Algorithm~\ref{alg3}, since they achieve the same level of privacy guarantees for preventing all possible sensitive inferences. Then, we compare the baseline method with our mechanism under usage mode 2 as shown in Algorithm~\ref{alg2}, to verify the effectiveness of our RLDP for protecting a specific subset of sensitive inferences.\\
\noindent\usenix{\secondrevise{\bf{Utility Advantage under Both Usage Modes:}}} Figure~\ref{acc_epsilon} shows the utility performance computed \revise{over the obfuscated sensor data generated by DEEProtect}. We can make the following important observations using Figure~\ref{acc_epsilon}: 1) DEEProtect achieves considerable advantage over the baseline approach with up to $8x$ improvement in utility. This validates the effectiveness of DEEProtect that not only provides rigorous privacy guarantees for protecting sensitive inferences, but also retains the utility of the perturbed data. \SP{Furthermore, we computed 95\% confidence interval for our accuracy results whose range is less than 4.83\%. These results further validate the efficacy of DEEProtect}; 2) DEEProtect achieves better utility performance under usage mode 2 which only considers specific sensitive inferences than that under usage mode 1 which considers the entire set of sensitive inferences; 3) As expected, at higher values of $\epsilon$, there is an improvement in utility but at the cost of degradation in privacy; 4) \ccsnew{Even at a moderate value of $\epsilon=5$ which is a typical privacy budget in LDP (similar values can also be found in Google's Rappor System~\cite{erlingsson2014rappor}), the authentication accuracy under usage mode 2 is $79.58\%$ as shown in Figure~\ref{feature_utility} which is close to the noise-free level.}  \ndsstwo{Note that the neighboring databases in LDP/RLDP may differ in all their possible tuples instead of differing in only one tuple as in DP. Thus, the LDP/RLDP guarantee is much stronger than DP guarantee making its typical privacy budget for balancing utility and privacy higher than that of DP.} \ndsstwo{Even though the typical privacy budget of our approach is around $\epsilon=5$, our system is still a significant improvement over existing SPPMs (such as ipShield~\cite{chakraborty2014ipshield}) that use heuristic obfuscation techniques and do not provide any provable guarantees.} \ndss{The accuracy of the sensitive inferences in Case Study 1 (activity mode detection) is close to a random guess at these $\epsilon$ values as will be discussed later. This observation further validates the effectiveness of our mechanisms.} \\
\indent To investigate the effectiveness of DEEProtect on real-world mobile applications, we plot the trade-off between the accuracy of making {useful inferences} versus the accuracy of making {sensitive inferences} for all the three case studies \unknown{(corresponding to usage mode 2 and thus Algorithm~\ref{alg2} is applied), as shown in Figure~\ref{balance123}}. \ccsfinal{We infer the useful/sensitive inference classifier as the ridge regression classifier in Eq.~\ref{krr_obj}, by utilizing the labeled training data corresponding to the useful/sensitive inference function. Note that our analysis are not restricted to the ridge regression algorithm and can be generalized to arbitrary inference methods.}\\
\noindent{{\bf{Utility-Privacy Tradeoffs for \ccs{Inference control}:}} {In Figure~\ref{balance1}}, we observe that 1) DEEProtect achieves good inference performance for behavior-based authentication (useful inference), while significantly deteriorating the activity mode detection (sensitive inference). For privacy budget {$\epsilon=5$ which is a typical privacy budget in LDP (similar values have also been used in Google's Rappor System~\cite{erlingsson2014rappor})}, the accuracy for inferring the {useful information} (authentication) is larger than $80\%$ while the accuracy for inferring sensitive information (activity modes) drops to roughly $5\%$ which is equivalent to random guessing ($1/20=5\%$). Therefore, DEEProtect can achieve practical privacy while only degrading utility performance by $10\%$. This provides an effective guide for users to choose a proper value of $\epsilon$ for real world applications; 2) higher levels of perturbation would degrade the inference performance for both the useful and sensitive information; 3) DEEProtect is effective for defending against sensitive inference attacks computed over sensor data. These observations demonstrate that DEEProtect works well in practice and returns an acceptable utility performance while satisfying provable privacy guarantees.}\\
\indent\secondrevise{Similarly, for the other two case studies, the accuracy of the sensitive inferences degrades at a much faster rate than that of the useful inferences when more noise is added (corresponding to a smaller privacy budget $\epsilon$), which further validates the effectiveness of DEEProtect.} 
For Case Study 2, in Figure~\ref{balance2}, when \usenix{$\epsilon = 5$}, DEEProtect achieves good performance for transportation detection (useful inference) \ccs{with accuracy higher than $78\%$}, while significantly degrading the identification of entered text (sensitive inference) \ccs{with accuracy less than $10\%$}. 
For Case Study 3, in Figure~\ref{balance3}, when \ccs{$\epsilon = 5$}, DEEProtect achieves good performance for speech translation (useful inference) \ccs{with accuracy higher than $79\%$}, while preventing the recognition of speaker identity (sensitive inference) \ccs{with accuracy less than $7\%$}.

\section{\ndss{Discussion, Limitations, Future Directions}}
\noindent\ndsstwo{\bf{Characteristics of Our Privacy Metrics:}} Depending on the mode the user chooses to operate in, DEEProtect can provide either the rigorous \usenix{LDP} guarantee that protects all sensitive inferences (usage mode 1) or our \usenix{RLDP} that protects a \ccs{specified} subset of inferences (usage mode 2). While RLDP provides weaker privacy guarantees than LDP, it comes with the advantage of stronger utility properties. When applying LDP/RLDP on mobile sensor data, the neighboring databases may differ in all their possible tuples (i.e., all sensor recordings across all timestamps within the same window), while the neighboring databases in traditional DP only differ in one tuple. Therefore, a proper privacy budget in LDP/RLDP for balancing privacy and utility is usually higher than that of DP \ndss{(similar privacy budgets have also been explored in~\cite{duchi2013local,erlingsson2014rappor,qin2016heavy}.)} \ndsstwo{According to the US chief scientific officer, the US Census in 2020 is planning to have all its datasets differentially private with $\epsilon=4$, which further validates the applicability of our privacy budget in real applications (recall Sections~\ref{per_exp}).}  
\ndssthree{Our privacy metrics also satisfy composition theorems and post-processing invariant properties (recall Theorems~\ref{seq},~\ref{paral},~\ref{ppip}).}

\noindent\ndsstwo{\bf Relationship between Useful and Sensitive Inferences:} \SP{For usage mode 1, the system configuration can specify as input a set of useful/authorized inferences and all other inferences are considered sensitive by default. For usage mode 1, the criterion to differentiate sensitive and useful inference is \emph{identical} to prior work on DP. For usage mode 2, the system configuration can specify a subset of inferences as being sensitive. For usage mode 2, we use the same criterion for specifying which inferences are acceptable as explored in previous research on permission granting~\cite{lin2014modeling}.}

\ndsstwo{Our perturbation mechanisms are general for any possible combination of useful and sensitive inferences. The parameter (privacy budget) selection process, similar to traditional DP frameworks, depends on the system's privacy-utility tradeoff requirement.} \ndsstwo{Our Case Study 3 considers speech-to-text translation as useful inference and speaker's identity recognition as sensitive inference. These two inferences are closely coupled since they both utilize MFCC and spectrogram features (recall Section~\ref{sec_approach}). The useful and sensitive inferences of other case studies in our experiments are also coupled. Experiments in Section~\ref{per_exp} demonstrate that our method can effectively balance the utility and privacy for these scenarios. The reason is that our RLDP aims to protect against a subset of sensitive inferences, instead of all the possible sensitive inferences computed over the sensor data. Therefore, even if those sensitive inferences are closely coupled with the useful inferences, the relaxed sensitivity $\Delta Q_{\mathrm{relax}}$ in Eq.~\ref{relaxed_sensitivity} is  still smaller than $\Delta Q$, making the perturbed sensor data usable in practice.} \ccsnew{Under situations where useful and sensitive inferences correspond to almost disjoint features, our RLDP mechanism can enhance the balance between utility and privacy.} \ndssthree{As the overlap in features corresponding to the useful and sensitive inferences increases, the performance of RLDP mechanism would decrease, approaching that of LDP mechanism.}

\noindent\ndsstwo{\bf Applicability and Generalizability:} \ccsnew{DEEProtect mechanisms provide obfuscated sensor data to applications, and are thus applicable to legacy apps as well.} {\ccsfinal{In contrast, the alternative paradigm of directly computing and sharing pre-defined utilities in a differentially private way is not compatible with legacy apps.}} Previous work in \cite{proserpio2014calibrating,shokri2015privacy} proposed to ignore several records in the database to bound sensitivity in DP for providing better privacy-utility tradeoffs. Our approach of using \ccs{autoencoder} techniques to extract authorized features is a conceptual improvement over these methods. While DEEProtect has been presented in the context of privacy-aware data sharing on mobile phones, the techniques developed are not restricted to the specific setting, and can be applied to other scenarios beyond mobile phones. We will make DEEProtect tool available to public as open source software.


\noindent\ndss{\bf Deployment Overhead:} \needcheck{We are working towards an implementation of the autoencoder on the Android platform using one of the existing packages~\cite{torch, caffe}.}
\ndss{Our current off-line training process requires a small amount of labeled data. In our experiments, we only required 20 users' data collected for 2 weeks, \ndss{which also validates the effectiveness of our training process.}} \ndss{On the mobile device, DEEProtect uses the pre-trained autoencoder model to transform data to features, performs feature perturbation and data reconstruction. These simple multiplication and addition operations performed on the mobile device are fairly lightweight.} 
\SP{\emph{In our experiments, we found that DEEProtect-on mode consumes 2.5\% more battery power, 3.8\% more CPU usage and 0.6MB more memory usage than the DEEProtect-off mode, which is an acceptable cost for daily usage.}} \SP{{Therefore, DEEProtect incurs fairly low overhead in off-line training and runtime resource consumption, indicating its high practicality.}} \needcheck{Any DP-oriented mechanism that adds noise to the data can risk the problem of breaking the integrity constraints of these sensors. In practice, since our framework targets simple sensors such as accelerometer, gyroscope, etc., our decoding process is unlikely to break the integrity constraints for these sensors.} \unknown{Therefore, smartphone apps can continue to use the perturbed sensor data generated by DEEProtect without any change to their existing interface.}

\noindent\ndsstwo{\bf Future Directions:} {While there is a non-trivial utility cost incurred by our mechanisms, DEEProtect (under both usage modes) significantly outperforms the baseline approach.} In addition, our utility performance can be further improved by 1) exploiting fine-grained correlation among sensors (e.g., among the three dimensions of accelerometer in Case Study 1) according to \cite{liu2016dependence}; 2) training customized models specific to each individual user that installs DEEProtect in phone; and 3) training models based on a larger amount of users' data as in \cite{erlingsson2014rappor}. Although our privacy guarantee is limited to segmented sensor data, we do take the temporal dynamics existing in mobile sensor stream into consideration according to a system-specified parameter of window size. Our perturbation mechanism can also be easily generalized to consider correlation across time windows, according to the composition properties of our privacy metrics (refer to \cite{duchi2013local} and Theorems~\ref{seq},\ref{paral}). An important component of our perturbation mechanism is the computation of the feature sensitivity in Algorithms~\ref{alg3},\ref{alg2}. Ways to accurately compute the sensitivity for arbitrary mobile applications, and deal with its possible underestimation would be a challenge we would like to address in the future. \ndssthree{Our repository of inferences in Table~\ref{tableknowledge} can be expanded as new useful/authorized and sensitive inferences are developed in the future.}

\section{conclusions}
In this paper, we propose DEEProtect, a general privacy-preserving framework for \ccs{inference control} of time-series sensor data on mobile devices. \revise{DEEProtect not only supports conventional LDP guarantees, but also provides a novel relaxed variant of LDP. We further propose effective perturbation mechanisms consisting of two key steps: 1) we explore novel autoencoder models to realize data minimization and only retain features relevant to the useful (authorized) inferences. This prevents the leakage of any information that is orthogonal to the useful inferences; and 2) to further enhance the privacy of sensitive inferences, we perturb the extracted features, using an effective obfuscation mechanism that ensures both conventional and relaxed LDP guarantees while simultaneously maintaining the utility of the data.} Finally, for reasons of compatibility with existing third-party apps, we reconstruct the sensor data, from the noisy features before sharing. Through theoretical analysis and extensive experiments over multiple real-world datasets, we demonstrate that compared to the state-of-the-art research, DEEProtect significantly improves the accuracy for supporting mobile sensing applications while providing provable privacy guarantees.

\section{Acknowledgement}
This work has been partly supported by faculty awards from Google, Intel and NSF. Changchang Liu is partly supported by IBM PhD Fellowship. We are very thankful to Brendan McMahan, Keith Bonawitz, Daniel Ramage, Nina Taft from Google, and Richard Chow from Intel for useful feedback and discussions.
{\bibliographystyle{IEEEtranS}
\bibliography{v3}}

\balance
\section{Appendix}
\begin{table*}[!t]
\renewcommand\arraystretch{1}
\centering
\caption{{\chang{Summary of Existing Inferences Computed over Mobile Sensor Data.}}}
\resizebox{\textwidth}{!}{
\begin{tabular}{c|c|c|c|c|c|c|c|c}
\hline
{} & {Accelerometer} & {Orientation} & {Gyroscope} & {{\tabincell{c}{Magnetic\\Field}} }  & Microphone & Camera & GPS& {{\tabincell{c}{Touch\\Screen}} } \\
\hline
{\tabincell{c}{Activity Mode Detection\\ \emph{attribute:~walking, still, etc.}}} & \cite{hemminki2013accelerometer,bao2004activity,reddy2010using} &  &  &  &  & & \\
\hline
{\tabincell{c}{Behavior-based Authentication\\ \emph{attribute:~authorized user or not}}}  & \cite{lee2015multi,mare2014zebra,conti2011swing,riva2012progressive,zhu2013sensec} & \cite{conti2011swing,lee2015multi} &  \cite{mare2014zebra,zhu2013sensec}& \cite{lee2015multi,zhu2013sensec} & \cite{riva2012progressive} &\cite{riva2012progressive} & &\cite{frank2013touchalytics,riva2012progressive}\\
\hline
{\tabincell{c}{Text Entered\\ \emph{attribute:~alphabets, digits}}} & \cite{xu2012taplogger,owusu2012accessory,marquardt2011sp,miluzzo2012tapprints} & \cite{xu2012taplogger,cai2011touchlogger} & \cite{miluzzo2012tapprints} &  &  & & &  \\
\hline
{\tabincell{c}{Speaker Identity Recognition\\ \emph{attribute:~user's identity}}} & \cite{liu2012cloud} & \cite{liu2012cloud}  & \cite{liu2012cloud}  & \cite{liu2012cloud}  & \cite{liu2012cloud}  & \cite{liu2012cloud,nirjon2013auditeur}  & & \\
\hline
{\tabincell{c}{Speech-to-text Translation\\ \emph{attribute:~speech to text}}} &  &  & \cite{michalevsky2014gyrophone} &  &  & \cite{lei2013accurate,zhou2013ibm} & & \\
\hline
{\tabincell{c}{Location \\ \emph{attribute:~home, work, public}}} &\cite{kim2010sensloc,han2012accomplice}&  &  & \cite{nirjon2013auditeur} &  & & \cite{brouwers2011detecting,kim2010sensloc} \\
\hline
{\tabincell{c}{Device Placement \\ \emph{attribute:~hand, ear, pocket, bag}}} &\cite{park2012online} & &  &  &  &  & & \\
\hline
{\tabincell{c}{Onscreen Taps \\ \emph{attribute:~location of apps on screen}}} &\cite{miluzzo2012tapprints} & &\cite{miluzzo2012tapprints}  &  &  &  & & \\
\hline
{\tabincell{c}{Stress \\ \emph{attribute:~stressful or not}}} & & &  &  & \cite{lu2012stresssense,chang2011ammon} &  & & \\
\hline
{\tabincell{c}{Emotion\\ \emph{attribute:~happy, sad, fear, anger}}}  & & &  &  & \cite{chang2011ammon,rachuri2010emotionsense} &  & & \\
\hline
{\tabincell{c}{Environment Monitor\\ \emph{attribute:~place virtualization}}}  & \cite{mohan2008nericell,PlaceRaider}& \cite{PlaceRaider}&  &  & \cite{azizyan2009surroundsense,mohan2008nericell,PlaceRaider} & \cite{azizyan2009surroundsense,mohan2008nericell,PlaceRaider} & & \\
\hline
\end{tabular}}
\label{tableknowledge}
\end{table*}

\subsection{Repository of Inferences}
To have a better understanding for current inference-based mobile sensing platforms, we investigate more than $70$ research papers published in relevant conferences and journals over the past $6$ years to form a knowledge repository in Table~\ref{tableknowledge}. This database captures a wide variety of inference categories over mobile sensor data including \emph{activity mode detection~\cite{hemminki2013accelerometer,bao2004activity,reddy2010using}, behavior-based authentication~\cite{frank2013touchalytics,lee2015multi,mare2014zebra,conti2011swing,riva2012progressive,zhu2013sensec}, text entered~\cite{xu2012taplogger,miluzzo2012tapprints,owusu2012accessory,liu2015good,cai2011touchlogger,marquardt2011sp}, speaker identity recognition~\cite{liu2012cloud,nirjon2013auditeur}, speech-to-text translation~\cite{lei2013accurate,michalevsky2014gyrophone,zhou2013ibm}, location~\cite{brouwers2011detecting,kim2010sensloc,nirjon2013auditeur,han2012accomplice} , device placement~\cite{park2012online}, onscreen taps~\cite{miluzzo2012tapprints}, stress~\cite{lu2012stresssense,chang2011ammon}, emotion~\cite{chang2011ammon,rachuri2010emotionsense} and environment monitoring~\cite{azizyan2009surroundsense,mohan2008nericell,PlaceRaider,tung2015echotag}}. While newer inferences are being made, we do not expect the database to change rapidly. Both the useful inferences and sensitive inferences for our systems are currently selected from the knowledge repository in Table~\ref{tableknowledge}.

\subsection{Sequential Composition Theorem of RLDP}
\noindent{\bf Proof for Theorem~\ref{seq}:} For $D_1$ and $D_2$, let $D_{1t}=D_1\cap D_{1t}$ and $D_{2t}=D_2\cap D_{2t}$. For any sequence $r$ of outcomes $r_t\in\mathbb{R}({A}_t)$, applying Definition~\ref{IPdef}, the probability of output $r$ from the sequence of ${A}_t(D_1)$ is 
\begin{equation}
\begin{aligned}
&\mathbb{P}(\widetilde{{A}}(D_1)=r)\\
=&\prod_t \mathbb{P}({A}_t(D_{1t})=r_t)\\
\le& \prod\limits_t \mathbb{P}({A}_t(D_{2t})=r_t)\times \prod\limits_t \exp(\frac{\epsilon_t}{\Delta D}\times |D_{1t}-D_{2t}|)\\
\le &\mathbb{P}(\widetilde{{A}}(D_2)=r)\times \exp(\frac{\sum_t \epsilon_t}{\Delta D} \times |D_1-D_2|)\\
\le& \mathbb{P}(\widetilde{{A}}(D_2)=r)\times \exp\left(\sum_t \epsilon_t\right)
\end{aligned}
\end{equation}

\subsection{Parallel Composition Theorem of RLDP}
\noindent {\bf Proof for Theorem~\ref{paral}:} For $D_1$ and $D_2$, let $D_{1t}=D_1\cup D_{1t}$ and $D_{2t}=D_2\cup D_{2t}$. For any sequence $r$ of outcomes $r_t\in\mathbb{R}({A}_t)$, applying Definition~\ref{IPdef}, the probability of output $r$ from the sequence of ${A}_t(D_1)$ is 
\begin{equation}
\begin{aligned}
&\mathbb{P}(\widetilde{{A}}(D_1)=r)
\\=&\prod_t \mathbb{P}({A}_t(D_{1t})=r_t)\\
\le& \prod\limits_t Pr({A}_t(D_{2t})=r_t)\times \prod\limits_t \exp(\frac{\epsilon_t}{\Delta D}\times |D_{2t}-D_{1t}|)\\
\le & \mathbb{P}(\widetilde{{A}}(D_2)=r)\times \exp(\frac{\max_t \epsilon_t}{\Delta D} \times |D_1-D_2|)\\
\le & \mathbb{P}(\widetilde{{A}}(D_2)=r)\times \exp(\max_t \epsilon_t)
\end{aligned}
\end{equation}

\subsection{Post-processing Invariant Property of RLDP}
\noindent {\ccs{\bf{Proof for Theorem~\ref{ppip}:}}} For $D_1$ and $D_2$, 
\begin{equation}
\begin{aligned}
&\mathbb{P}(f(A(D_1))=s)\\
=&\mathbb{P}(A(D_1)=r)\\
\le& \exp(\epsilon)\mathbb{P}(A(D_2)=r)\\
=&\exp(\epsilon)\mathbb{P}(f(A(D_2))=s)
\end{aligned}
\end{equation}

\subsection{Privacy Guarantees of the Baseline Approach}
For the baseline approach that adds Laplace noise with parameter $\lambda=\dim({\bm{x}}_t)\Delta Q/\epsilon$ to the sensor data, we have 
\begin{equation}
\begin{aligned}
&\max\limits_{{\bm{x}}_{t1},{\bm{x}}_{t2},o}\frac{\mathbb{P}({A}({\bm{x}}_{t1})={o})}{\mathbb{P}({A}({\bm{x}}_{t2})={o})}\\
\le& \max\limits_{{\bm{x}}_{t1},{\bm{x}}_{t2},o}\frac{Lap({o}-Q({\bm{x}}_{t1}))}{Lap({o}-Q({\bm{x}}_{t2}))}\\
\le& \max\limits_{{\bm{x}}_{t1},{\bm{x}}_{t2},o}\frac{\exp\left(\frac{\epsilon({o}-Q({\bm{x}}_{t1})))}{\dim({\bm{x}}_t)\Delta Q}\right)}{\exp\left(\frac{\epsilon({o}-Q({\bm{x}}_{t2})))}{\dim({\bm{x}}_t)\Delta Q}\right)}\\
\le&\exp{(\epsilon)}
\end{aligned}
\end{equation}

The above result holds since $\max\limits_{{\bm{x}}_{t1},{\bm{x}}_{t2}}\|Q({\bm{x}}_{t1})-Q({\bm{x}}_{t2})\|=\dim({\bm{x}}_t)\Delta Q$ (according to the composition properties of DP).

\subsection{Privacy Guarantees of DEEProtect Under Usage Mode 1}
\noindent{\bf Proof for Theorem~\ref{proofprivacy}:} {First, we have $\Delta {\bm{f}}_t\le \sqrt{\dim({\bm{f}}_t)}\Delta_2 {\bm{f}}_t \le \sqrt{\dim({\bm{f}}_t)}\Delta_2 {\bm{x}}_t \le \sqrt{\dim({\bm{f}}_t)}\Delta {\bm{x}}_t \le \sqrt{\dim({\bm{f}}_t)}\dim({\bm{x}}_t)\Delta Q$, where $\Delta_2(\cdot)$ represents norm-2 distance. Thus, by setting $\lambda=\frac{\sqrt{\dim({\bm{f}}_t)}\dim({\bm{x}}_t)\Delta Q}{\epsilon}$, we have }
\begin{equation}
\begin{aligned}
&\max\limits_{{\bm{f}}_{t1},{\bm{f}}_{t2},o}\frac{\mathbb{P}({A}({\bm{f}}_{t1})={o})}{\mathbb{P}({A}({\bm{f}}_{t2})={o})}\\
=&\max\limits_{{\bm{f}}_{t1},{\bm{f}}_{t2},o}\frac{Lap({o}-Q({\bm{f}}_{t1}))}{Lap({o}-Q({\bm{f}}_{t2}))}\\
\le& \max\limits_{{\bm{f}}_{t1},{\bm{f}}_{t2},o}\frac{\exp\left(\frac{\epsilon({o}-Q({\bm{f}}_{t1})))}{\sqrt{\dim({\bm{f}}_t)}\dim({\bm{f}}_t)\Delta Q}\right)}{\exp\left(\frac{\epsilon({o}-Q({\bm{f}}_{t2})))}{\sqrt{\dim({\bm{f}}_t)}\dim({\bm{f}}_t)\Delta Q}\right)}\\
\le&\exp{(\epsilon)}
\end{aligned}
\end{equation}

{Then, according to the post-processing invariance property of LDP, we have $\max\limits_{{\bm{x}}_{t1},{\bm{x}}_{t2},o}\frac{\mathbb{P}({A}({\bm{x}}_{t1})={o})}{\mathbb{P}({A}({\bm{x}}_{t2})={o})}
\le\exp{(\epsilon)}$.}
\subsection{Privacy Guarantees of DEEProtect Under Usage Mode 2}
\noindent{\bf Proof for Theorem~\ref{proofprivacy2}:} {By setting $\lambda=\frac{\sqrt{\dim({\bm{f}}_t)}\dim({\bm{x}}_t)\Delta Q_{\mathrm{relaxed}}}{\epsilon}$, we have}

\begin{equation}
\begin{aligned}
&\max\limits_{{\bm{f}}_{t1},{\bm{f}}_{t2},o}\frac{\mathbb{P}({A}({\bm{f}}_{t1})={o})}{\mathbb{P}({A}({\bm{f}}_{t2})={o})}\\
\le& \max\limits_{{\bm{f}}_{t1},{\bm{f}}_{t2},o}\frac{Lap({o}-Q({\bm{f}}_{t1}))}{Lap({o}-Q({\bm{f}}_{t2}))}
\\
\le& \max\limits_{{\bm{f}}_{t1},{\bm{f}}_{t2},o}\frac{\exp\left(\frac{\epsilon({o}-Q({\bm{f}}_{t1})))}{\sqrt{\dim({\bm{f}}_t)}\dim({\bm{f}}_t)\Delta Q_{\mathrm{relaxed}}}\right)}{\exp\left(\frac{\epsilon({o}-Q({\bm{f}}_{t2})))}{\sqrt{\dim({\bm{f}}_t)}\dim({\bm{f}}_t)\Delta Q_{\mathrm{relaxed}}}\right)}\\
\le& \exp{(\epsilon)}
\end{aligned}
\end{equation} 

{Therefore, according to the post-processing invariance property of RLDP, we have $\max\limits_{{\bm{x}}_{t1},{\bm{x}}_{t2},o}\frac{\mathbb{P}({A}({\bm{x}}_{t1})={o})}{\mathbb{P}({A}({\bm{x}}_{t2})={o})}
\le\exp{(\epsilon)}$.}

\subsection{Utility Analysis of DEEProtect Under Usage Mode 1}
\noindent{\bf Proof for Theorem~\ref{the_error}:}  First, we derive the variance of a mechanism ${A}(\cdot)$ for DEEProtect under usage mode 1 as
\begin{equation}
\begin{aligned}
Var({A}({\bm{x}}_{t}))=&\sum\limits_{i=1}^{\dim({\bm{f}}_{t})}\frac{Var({\bm{f}}_{t}^\prime)}{{\dim({\bm{x}}_{t})}^2}\\
\le&\frac{{{\dim({\bm{f}}_{t})}}^2(\Delta {\bm{f}}_{t}(i))^2}{{\dim({\bm{x}}_{t})}^2\epsilon^2}\\
\le&\frac{{\dim({\bm{f}}_{t})}^4(\Delta Q)^2}{{\dim({\bm{x}}_{t})}^2\epsilon^2}\\
\le&\frac{{\dim({\bm{f}}_{t})}^2(\Delta Q)^2}{{\dim({\bm{x}}_{t})}^2\epsilon^2}\\
\end{aligned}
\end{equation}
Its expected error is
\begin{equation}
\begin{aligned}
&\mathbb{E}[|{A}({\bm{x}}_t)-{\bm{x}}_t|] \\
\le& \mathbb{E}[\|{A}({\bm{x}}_t)-\mathbb{E}[{A}({\bm{x}}_t)]\|_1]+\mathbb{E}[\|{A}({\bm{x}}_t)-{\bm{x}}_t\|_1]\\
=&Re\_{error}({A}({\bm{x}}_t))+\sqrt{\mathbb{E}[\|{A}({\bm{x}}_t)-{\bm{x}}_t\|_2^2]}\\
=&Re\_{error}({A}({\bm{x}}_t))+\sqrt{Var({A}({\bm{x}}_t))}\\
\end{aligned}
\end{equation}
Note that our \ccs{autoencoder} based data minimization mechanism would result in a negligible reconstruction error, i.e., $Re\_{error}(A({\bm{x}}_t))$ is much lower than the expected error caused by adding noise. Therefore, we approximate the utility performance of DEEProtect as 
\begin{equation}
Error({A})\approx \frac{\dim({\bm{f}}_t)\Delta Q }{\dim({\bm{x}}_t)\epsilon}
\end{equation}
Similar results can also be found in~\cite{rastogi2010differentially}. The expected error for the baseline approach can also be obtained as 
\begin{equation}
\begin{aligned}
\mathbb{E}[\|LPM({\bm{x}}_t)-{\bm{x}}_t\|_1]
=&\mathbb{E}[\|(Lap(\Delta Q/\epsilon))\|_1]\\
=&\Delta Q/\epsilon
\end{aligned}
\end{equation}

Therefore, DEEProtect under usage mode 1 reduces the expected error of the baseline approach with a factor of $\dim({{\bm{x}}_t})/\dim({{\bm{f}}_t})$.

\subsection{Utility Analysis of DEEProtect Under Usage Mode 2}
\noindent{\bf{Proof for Theorem~\ref{the_error2}:}} First, we derive the variance of a mechanism ${A}(\cdot)$ for DEEProtect under usage mode 2 as 
\begin{equation}
\begin{aligned}
&\mathbb{E}[|{A}({\bm{x}}_t)-{\bm{x}}_t|] \\
\le &\mathbb{E}[\|{A}({\bm{x}}_t)-\mathbb{E}[{A}({\bm{x}}_t)]\|_1]+\mathbb{E}[\|{A}({\bm{x}}_t)-{\bm{x}}_t\|_1]\\
=&Re\_{error}({A}({\bm{x}}_t))+\sqrt{\mathbb{E}[\|{A}({\bm{x}}_t)-{\bm{x}}_t\|_2^2]}\\
=&Re\_{error}({A}({\bm{x}}_t))+\sqrt{Var({A}({\bm{x}}_t))}
\end{aligned}
\end{equation}

Similarly, we have 
\begin{equation}
\begin{aligned}
Error({A})\approx \frac{\dim({{\bm{f}}_t})\Delta Q_{\mathrm{relaxed}}}{\dim({\bm{x}}_t)\epsilon}
\end{aligned}
\end{equation} 

Therefore, we know that DEEProtect under usage mode 2 reduces the expected error of the baseline approach with a factor of $\frac{\dim({\bm{x}}_t)\Delta Q}{\dim({\bm{f}}_t)\Delta {Q_{\mathrm{relaxed}}}}$. 

\subsection{Existing Autoencoder Models}
The mathematical formulations of existing autoencoder models~\cite{hinton2006reducing,lee2008sparse} (expanding Eq.~\ref{cost}) is 

\begin{equation}
\begin{aligned}
\min J_0({\bm{W}},{\bm{b}}_e,{\bm{b}}_d,{\bm{W}}^\prime)=&
\min\sum\nolimits_{t=1}^{T/N_w} L({\bm{x}}_t,D_{ec}(E_{nc}({\bm{x}}_t)))\\
+&\lambda\sum\nolimits_{t,i}{\bm{W}}_{t,i}^2+\delta\sum\nolimits_{i=1}^{d_{\bm{f}}}KL(\rho||\hat\rho_i)
\end{aligned}
\end{equation}

 \usenix{where the encoder function $E_{nc}(\cdot)$ maps the input data ${\bm{x}}_t\in\mathbb{R}^{d_{{\bm{x}}_t}\times 1}$ to the hidden units (features) ${\bm{f}}_t\in\mathbb{R}^{d_{{\bm{f}}_t}\times 1}$ according to}

\begin{equation}
{\bm{f}}_t=E_{nc}({\bm{x}}_t)=g_{enc}({\bm{W}}{\bm{x}}_t+{\bm{b}}_e)
\end{equation}
and the decoder function $D_{ec}(\cdot)$ maps the outputs of the hidden units (features) back to the original input space according to

\begin{equation}
\widetilde{\bm{x}}_t=D_{ec}({\bm{f}}_t)=g_{dec}({\bm{W}}^\prime{\bm{f}}_t+{\bm{b}}_d)
\end{equation}

 Similar to most existing \ccs{autoencoder} models~\cite{hinton2006reducing, lee2008sparse, vincent2008extracting}, we choose the sigmoid function for $g_{enc}(\cdot), g_{dec}(\cdot)$, and set ${\bm{W}}^\prime={\bm{W}}^T$. $L({\bm{u}},{\bm{v}})$ is a loss function, typically the square loss $L({\bm{u}},{\bm{v}})=\|{\bm{u}-{\bm{v}}}\|^2$. The second term is a weight decay term that helps prevent overfitting~\cite{hinton2006reducing}. The third term is a sparsity constraint 

\begin{equation}
\sum_{i=1}^{d_{\bm{f}}}KL(\rho||\hat \rho_i)
\end{equation}
with parameter $\rho$, where 

\begin{equation}
\hat\rho_i=\frac{1}{n}\sum_{t=1}^{T/N_w} f_i({\bm{x}}_t)
\end{equation} 

$KL(\rho||\hat\rho_i)$ is the KL divergence between two Bernoulli random variables with mean $\rho$ and $\hat \rho_i$ respectively.

\subsection{Model Learning in Autoencoder-based Data Minimization}
To solve the optimization problem in Eq.~\ref{final_obj}, we first initialize each parameter ${\bm{W}},{\bm{b}}_e,{\bm{b}}_d,{\bm{W}}^\prime,{\bm{c}}$ to a small random value near zero. Then we apply SGD~\cite{bottou1991stochastic} for iterative optimization. We perform the feedforward pass in Eq.~\ref{final_obj} and evaluate 
	\begin{equation}
	{\bm{z}}_t^e={\bm{Wd}}_t+{\bm{b}}_e,
	{\bm{z}}_t^{d}={\bm{W}}^Ts_e({\bm{z}}_t^e)+{\bm{b}}_d
	\end{equation}  
	
	Next, we compute the error terms $\bm{\delta}_t^d, {\bm{\delta}}_t^e$ as 
	\begin{equation}
	\begin{aligned}
	{\bm{\delta}}_t^d&=\frac{\partial L(\widetilde{\bm{d}}_t, d(e({\bm{d}}_t)))}{\partial {\bm{z}}_t^d}=-(\frac{\widetilde{{\bm{d}}}_t}{s_d({\bm{z}}_t^d)}+\frac{1-\widetilde{{\bm{d}}}_t}{1-s_d({\bm{z}}_t^d)})s_d^\prime({\bm{z}}_t^d)\\
	{\bm{\delta}}_t^e&=({\bm{W}}{\bm{\delta}}_t^d+\delta({\rho}/{\widetilde{\rho}_t}+(1-\rho)/(1-\widetilde{\rho}_t)))s_{t}^\prime({\bm{z}}_t^e)
	\end{aligned}
\end{equation}	
	
	 Finally, the gradient descents as partial derivatives can be computed as 
	 \begin{equation}
	 \begin{aligned}
	\frac{\partial J({\bm{W}},{\bm{b}}_e,{\bm{b}}_d,{\bm{W}}^\prime,{\bm{c}})}{\partial {\bm{W}}}=&\sum_{t=1}^T g_{dec}({\bm{x}}_t){\bm{\delta}}_t^{dT}+{\bm{\delta}}_t^{e}{\bm{x}}_t^T\\
	 \frac{\partial J({\bm{W}},{\bm{b}}_e,{\bm{b}}_d,{\bm{W}}^\prime,{\bm{c}})}{\partial {\bm{b}}_e}=&\sum_{t=1}^T {\bm{\delta}}_t^d\\
	  \frac{\partial J({\bm{W}},{\bm{b}}_e,{\bm{b}}_d,{\bm{W}}^\prime,{\bm{c}})}{\partial {\bm{b}}_d}=&\sum_{t=1}^T {\bm{\delta}}_t^e\\
	  \frac{\partial J({\bm{W}},{\bm{b}}_e,{\bm{b}}_d,{\bm{W}}^\prime,{\bm{c}})}{\partial {\bm{c}}}=&{\bm{c}^T}-\sum_{t=1}^T {\bm{x}}_t
	  \end{aligned}
	  \end{equation}
	   
	   {Next, we modify $\bm{W}$ to satisfy the orthogonality constraint in Eq.~\ref{final_obj}.} Our method to overcome this non-convex constraint follows the rigorous analysis in \cite{nam2014local} by adopting a simple operation, 
	   \begin{equation}
	   {\bm{W}}:=({\bm{WW}}^T)^{-\frac{1}{2}}\bm{W}
	   \end{equation}
	   
	    which sets the singular values of $\bm{W}$ to be all ones. 
	
\subsection{Model Stacking in Autoencoder-based Data Minimization}
 \unknown{Although Algorithm~\ref{alg1} is effective for solving Eq.~\ref{final_obj}, the learnt result heavily relies on the initial seeds. Therefore, we use multiple hidden layers to stack the model in order to achieve more stable performance.} For the first layer in the \ccs{autoencoder} model, we find the optimal layer by minimizing the objective function in Eq.~\ref{final_obj}. The representations learned by the first layer are then used as the input of the second layer, and so on so forth. \needcheck{Using a stacked multi-layer autoencoder, we can learn more stable and finer-grained features to better represent the mobile sensor data.}

\end{document}